\documentclass[prd,showpacs,preprintnumbers,amsmath,amssymb,nofootinbib,aps]{revtex4}

\usepackage{dcolumn}
\usepackage{bm}
\usepackage{amsfonts}
\usepackage{mathrsfs}
\usepackage{graphicx}

\begin{document}

\newcommand{\hhat}[1]{\hat {\hat{#1}}}
\newcommand{\pslash}[1]{#1\llap{\sl/}}
\newcommand{\kslash}[1]{\rlap{\sl/}#1}
\newcommand{\lab}[1]{}
\newcommand{\iref}[2]{}
\newcommand{\sto}[1]{\begin{center} \textit{#1} \end{center}}
\newcommand{\rf}[1]{{\color{blue}[\textit{#1}]}}
\newcommand{\eml}[1]{#1}
\newcommand{\el}[1]{\label{#1}}
\newcommand{\er}[1]{Eq.\eqref{#1}}
\newcommand{\df}[1]{\textbf{#1}}
\newcommand{\mdf}[1]{\pmb{#1}}
\newcommand{\ft}[1]{\footnote{#1}}
\newcommand{\n}[1]{$#1$}
\newcommand{\fals}[1]{$^\times$ #1}
\newcommand{\new}{{\color{red}$^{NEW}$ }}
\newcommand{\ci}[1]{}
\newcommand{\de}[1]{{\color{green}\underline{#1}}}
\newcommand{\ke}{\rangle}
\newcommand{\br}{\langle}
\newcommand{\lb}{\left(}
\newcommand{\rb}{\right)}
\newcommand{\blb}{\Big(}
\newcommand{\brb}{\Big)}
\newcommand{\nn}{\nonumber \\}
\newcommand{\p}{\partial}
\newcommand{\pd}[1]{\frac {\partial} {\partial #1}}
\newcommand{\cd}{\nabla}
\newcommand{\cc}{$>$}
\newcommand{\ba}{\begin{eqnarray}}
\newcommand{\ea}{\end{eqnarray}}
\newcommand{\be}{\begin{equation}}
\newcommand{\ee}{\end{equation}}
\newcommand{\bay}[1]{\left(\begin{array}{#1}}
\newcommand{\eay}{\end{array}\right)}
\newcommand{\eg}{\textit{e.g.} }
\newcommand{\ie}{\textit{i.e.}, }
\newcommand{\iv}[1]{{#1}^{-1}}
\newcommand{\st}[1]{|#1\ke}
\newcommand{\at}[1]{{\Big|}_{#1}}
\newcommand{\zt}[1]{\texttt{#1}}
\def\xa{{\alpha}}
\def\xA{{\Alpha}}
\def\xb{{\beta}}
\def\xB{{\Beta}}
\def\xd{{\delta}}
\def\xD{{\Delta}}
\def\xe{{\epsilon}}
\def\xE{{\Epsilon}}
\def\xve{{\varepsilon}}
\def\xg{{\gamma}}
\def\xG{{\Gamma}}
\def\xk{{\kappa}}
\def\xK{{\Kappa}}
\def\xl{{\lambda}}
\def\xL{{\Lambda}}
\def\xo{{\omega}}
\def\xO{{\Omega}}
\def\xvp{{\varphi}}
\def\xs{{\sigma}}
\def\xS{{\Sigma}}
\def\xt{{\theta}}
\def\xvt{{\vartheta}}
\def\xT{{\Theta}}
\def \Tr {{\rm Tr}}
\def\CA{{\cal A}}
\def\CC{{\cal C}}
\def\CD{{\cal D}}
\def\CE{{\cal E}}
\def\CF{{\cal F}}
\def\CH{{\cal H}}
\def\CJ{{\cal J}}
\def\CK{{\cal K}}
\def\CL{{\cal L}}
\def\CM{{\cal M}}
\def\CN{{\cal N}}
\def\CO{{\cal O}}
\def\CP{{\cal P}}
\def\CQ{{\cal Q}}
\def\CR{{\cal R}}
\def\CS{{\cal S}}
\def\CT{{\cal T}}
\def\CV{{\cal V}}
\def\CW{{\cal W}}
\def\CY{{\cal Y}}
\def\BC{\mathbb{C}}
\def\BR{\mathbb{R}}
\def\BZ{\mathbb{Z}}
\def\sA{\mathscr{A}}
\def\sB{\mathscr{B}}
\def\sF{\mathscr{F}}
\def\sG{\mathscr{G}}
\def\sH{\mathscr{H}}
\def\sJ{\mathscr{J}}
\def\sL{\mathscr{L}}
\def\sM{\mathscr{M}}
\def\sN{\mathscr{N}}
\def\sO{\mathscr{O}}
\def\sP{\mathscr{P}}
\def\sR{\mathscr{R}}
\def\sQ{\mathscr{Q}}
\def\sS{\mathscr{S}}
\def\sX{\mathscr{X}}

\def\slz{SL(2,Z)}
\def\slr{$SL(2,R)\times SL(2,R)$ }
\def\ads{${AdS}_5\times {S}^5$ }
\def\adst{${AdS}_3$ }
\def\sun{SU(N)}
\def\ad#1#2{{\frac \delta {\delta\sigma^{#1}} (#2)}}
\def\bqf{\bar Q_{\bar f}}
\def\nf{N_f}
\def\sunf{SU(N_f)}

\def\dcirc{{^\circ_\circ}}

\author{Xing Huang}
\email{huangx@uwm.edu}
\author{Leonard Parker}
\email{leonard@uwm.edu}
\affiliation{Physics Department, University of Wisconsin-Milwaukee,
P.O.Box 413, Milwaukee, Wisconsin USA 53201}

\title{Clarifying Some Remaining Questions in the Anomaly Puzzle}
\date{June 17, 2010}

\begin{abstract}
We discuss several points that may help to clarify some questions that remain about the anomaly puzzle in supersymmetric theories. In particular, we consider a general ${\cal N}=1$ supersymmetric Yang-Mills theory. The anomaly puzzle concerns the question of whether there is a consistent way to put the $R$-current and the stress tensor in a single supercurrent, even though in the classical theory they are in the same supermultiplet. As is well known, the classically conserved supercurrent bifurcates into two supercurrents having different anomalies in the quantum regime. The most interesting result we obtain is an explicit expression for the lowest component of one of the two supercurrents in 4-dimensional spacetime, namely the supercurrent that has the energy-momentum tensor as one of its components. This expression for the lowest component is an energy-dependent linear combination of two chiral currents, which itself does not correspond to a classically conserved chiral current. The lowest component of the other supercurrent, namely, the $R$-current, satisfies the Adler-Bardeen theorem. The lowest component of the first supercurrent has an anomaly that we show is consistent with the anomaly of the trace of the energy-momentum tensor. Therefore, we conclude that there is no consistent way to put the $R$-current and the stress tensor in a single supercurrent in the quantized theory. We also discuss and try to clarify some technical points in the derivations of the two-supercurrents in the literature. These latter points concern the significance of infrared contributions to the NSVZ $\beta$-function and the role of the equations of motion in deriving the two supercurrents.

\pacs{11.10.Hi, 11.30.Pb, 11.30.Rd}

\end{abstract}

\maketitle

\section{Introduction}
The anomaly puzzle in $\CN = 1$ supersymmetric gauge theories is well known.  
A real superfield, $ \CJ_{\mu}$, called the supercurrent
can be constructed \cite{Ferrara:1974pz} and is classically conserved. The lowest component of this superfield is the R-current.  The other components of $ \CJ_{\mu}$
are related to the supersymmetry current $J_{\xa\mu}$ (where $\xa$ is a two-component spinor index that labels the generators of the supersymmetry) and the stress tensor $\vartheta_{\mu\nu}$ through linear transformations. This construction is related to the fact that these symmetries are elements of the superconformal algebra. 

The anomaly puzzle arises as follows.
In an $\CN = 1$ SYM (supersymmetric Yang-Mills) theory, the R-symmetry, which is just a chiral $U(1)$ symmetry (denoted later as $U(1)_R$) has an anomaly.
This chiral anomaly is proportional to the topological invariant,
$F^{\mu\nu} {\tilde F}_{\mu\nu}$,  and can be expressed in an operator equation. One can try to generalize the operator equation
of this anomaly of the $R$-symmetry to a supersymmetric form involving $ \CJ_{\mu}$ \cite{Clark:1979te,Piguet:1981mu,Curtright:1977cg,Abbott:1977in,Lukierski:1977dr,Inagaki:1977he}.
However, this attempt led to an apparent contradiction. On the one hand, the anomaly of R-symmetry is known to be exactly of one-loop order because of the Adler-Bardeen theorem \cite{Adler:1969er,Jones:1982zf}. On the other hand, the trace of the stress tensor, which is another component of $D^\xa \CJ_{\xa \dot \xa}$ should be proportional to the $\xb$-function (because the trace is a measure of the breaking of scale invariance). These two components of $D^\xa \CJ_{\xa \dot \xa}$ should be proportional to the same factor, which would seem to imply that the $\xb$-function is exactly of one loop order. 
However, explicit perturbative calculations show that there are higher order corrections to the $\xb$-function \cite{Avdeev:1980bh}. Note that there are some subtleties about this formulation of the anomaly puzzle, which we shall discuss in more detail later.

In Grisaru et al, \cite{Grisaru:1985yk,Grisaru:1985ik}, a solution to the anomaly puzzle is given by showing that there are actually two different supercurrents $ \CJ_{\mu}$.
Let us call those two different supercurrents in 4-dimensional spacetime, $ \CJ^{(1)}{}_{\mu}$ and $ \CJ^{(2)}{}_{\mu}$.  They are the same classically (meaning at tree level). One of them, $ \CJ^{(1)}{}_{\mu}$, has the R-current as its lowest component, but the higher components are no longer the supersymmetry current and stress tensor. The anomalous non-conservation of this supercurrent is proportional to the one-loop $\xb$-function. The other supercurrent, $ \CJ^{(2)}{}_{\mu}$, has the supersymmetry current and stress tensor as its components and has an anomaly proportional the exact $\xb$-function. 
In Ensign et al \cite{Ensign:1987wy}, they consider $\CN = 1$ supersymmetric gauge theories including matter fields and extend the construction done in \cite{Grisaru:1985yk,Grisaru:1985ik} of the two supercurrents to the case that includes matter. 

In Shifman and Vainshtein \cite{Shifman:1986zi}, they argued for a different solution to the anomaly puzzle. They considered the coefficient in front of the $W^2$ term in the Wilsonian effective action and showed that its running, obtained by integrating out higher momentum modes, is only of one-loop order. (This result is in agreement with the nonrenormalization theorem, as formulated for example in \cite{Seiberg:1993vc}.)  
On the other hand, the physical coupling constant, defined from the physical amplitudes one measures in experiments, includes higher-order contributions.  Motivated by this, they proposed an operator anomaly equation with a one-loop coefficient. 
They showed that the $\xb$-function of the physical coupling is obtained when one takes the matrix element of their anomaly equation. 
 They assumed that their single operator supercurrent contains as its lowest component the R-current and as another component the stress tensor. 
However, we show below that this is not the case, by proving that the supercurrent, $ \CJ^{(2)}{}_{\mu}$, having the energy-momentum tensor as one of its components does not, in fact, have the R-current as its lowest component. 
To avoid any ambiguity, we mention that we are using the term $R$-current (and $R$-symmetry) to describe the $U(1)$ current (denoted by $R_\mu$) that transforms the gaugino $\xl$,
the matter scalar $A$ and the matter spinor $\psi$ according to the charge ratios of $1:\frac 2 3: - \frac 1 3$. 
We find that $\CJ^{(2)}{}_{\mu}$ has as its lowest component a current which is an energy-dependent linear combination of the R-current and the Konishi current. This linear combination, which we refer to as $R'_{\mu}$, is not a chiral current.  This favors the idea of two supercurrents proposed in \cite{Grisaru:1985yk, Grisaru:1985ik, Ensign:1987wy}. The explicit expression for the lowest component of this supercurrent had not been written earlier to our knowledge.
%
%
%
%
%

The anomaly equation for
SYM with matter fields, as given in \cite{Shifman:1986zi}, has a term $\xg \bar D^2 (\bar \Phi e^V \Phi)$
(where $\Phi$ is a chiral superfield)
that is responsible for the anomalous dimensions of the matter fields.  
\eml{This term does not appear in \cite{Ensign:1987wy} because they assume that external fields are on-shell.} 
As we shall see, it is the existence of this term that implies that the lowest component of $R'_\mu$ is not the $R$-current but a mixing (with coupling constant dependent coefficients) of the $R$-current and the Konishi current. We perform an explicit calculation, which is not in the literature, to obtain the mixing. \eml{We perform the calculation using both component fields and the supersymmetric
background field method.} The results we obtain from either method agree and give the $\xg \bar D^2 (\bar \Phi e^V \Phi)$ term.  

The Adler-Bardeen theorem implies that the chiral current $R_\mu$ has an anomaly of one-loop order, but does not imply that $R'_\mu$
should have a one-loop anomaly.
We also find that the difference between $R'_\mu$ and $R_\mu$ is manifest in a very clear way at the infrared fixed point, where $R'_\mu$ becomes a non-anomalous symmetry current that is a linear combination of $R_\mu$ and the Konishi current. \eml{We use the supersymmetric QCD model as an example to show this behavior of the operator $R_\mu'$.}


Although the two-supercurrent scenario appears to be
the correct solution to the anomaly puzzle, there
are some technical issues in their construction that we discuss and attempt to clarify. In \cite{Grisaru:1985ik}, the equations of motion (EoM) are applied with the assumption that they vanish (up to contact terms). However, if one uses the expectation values of the various operators, as given in \cite{Grisaru:1985ik}, then the EoM would seem to have nonvanishing expectation values.  We show that this apparent inconsistency is resolved when one takes into account the non-local contributions.
After that, the expectation values of the bare operators are consistent with the application of the EoM. 
In particular, the expectation value, $\br\cd^\xa \CJ_{\xa \dot \xa}\ke$, of the unrenormalized operator $\cd^\xa \CJ_{\xa \dot \xa}$ vanishes as required to by the EoM. More explicitly, the non-local contribution to $\br\cd^\xa \CJ_{\xa \dot \xa}\ke$ is opposite in sign to the local contribution, which is proportional to an $\xe$ dimensional operator \ft{The
calculation is performed using dimensional
reduction and the dimension is $4-\xe$ with
$\xe>0$.}, and the two contributions add up to zero in the limit that $\xe\rightarrow 0$, i.e., in 4 dimensions. As a result, 
$\br\cd^\xa \CJ_{\xa \dot \xa}\ke$ does vanish.  Then, when we use the renormalization procedure of \cite{Grisaru:1985yk,Grisaru:1985ik,Ensign:1987wy}, in which the contribution proportional to an $\xe$ dimensional operator is removed by renormalization, the non-local contribution indeed gives the correct one-loop anomaly. This correct one-loop anomaly was obtained in \cite{Grisaru:1985yk,Grisaru:1985ik,Ensign:1987wy}. 
They did not explicitly discuss the role played by the non-local contributions in their derivation, so the discussion of those terms here may help clarify the consistency of the construction of the two supercurrents.

Finally, we comment on the question of whether the higher-order terms in the $\xb$-function are the result of contributions coming from infrared modes of the fields. In \cite{Shifman:1986zi}, they show that the higher-order terms in the $\xb$-function come from the infrared modes. A different way of obtaining the same $\xb$-function is given in \cite{ArkaniHamed:1997mj}. In the latter method, the coupling constant receives its higher-order corrections from the Jacobian appearing when one rescales the measure \cite{Fujikawa:1979ay,Fujikawa:1980eg}, and as they mention in \cite{ArkaniHamed:1997mj}, the method does not appear to depend on the infrared modes. By changing the UV cutoff in the Wilson effective action, we show that the momentum modes above any arbitrary finite non-zero scale do not give a significant contribution to the Jacobian from which the multi-loop corrections to the $\beta$-function are obtained.  This shows that the method used by \cite{ArkaniHamed:1997mj} does indeed depend on the infrared modes.

In section II, we review some basic ideas about the supercurrent and the anomaly puzzle. The supercurrent is discussed in more detail in the appendix. In section III, possible solutions to the anomaly puzzle in the literature are reviewed and remaining problems are discussed. In section IV, \eml{we perform an explicit calculation to show that the operator $R_\mu'$ in the same supermultiplet as the supersymmetry
current has exactly the properties of what the anomaly equation in \cite{Shifman:1986zi}
predicts but it generates a $U(1)$ transformation different from the $R$-symmetry. As a result,
this superfield should be identified as $\CJ^{(2)}{}_{\mu}$
and not as $\CJ^{(1)}{}_{\mu}$ (in the notation defined above). First we do the calculation using component fields. Then in section IVA, we obtain the same result using the supersymmetric background field method. In section IVB, we analyze the properties
of the current $R_\mu'$ at the non-trivial infrared fixed point of supersymmetric QCD. We
show that $R_\mu'$ does have the charge ratios to be a non-anomalous current and thus corresponds to
a true symmetry at the fixed point, as it should.} 
In section V, we discuss the role of non-local terms in obtaining the expectation value of the equation of motion and show how such terms enter into the construction of the two supercurrents.
In section VI, we show that the calculations
of the $\xb$-function done by Shifman and Vainshtein \cite{Shifman:1986zi} and by Arkani-Hamed and Murayama \cite{ArkaniHamed:1997mj} both depend on the infrared modes. 

\section{Review of Anomaly Puzzle}
First of all, let us review the basics of the supercurrent and the anomaly puzzle. $\CN = 1$ gauge theory is described by a Lagrangian, 
\ba \el{APN1SYM:L}\CL & = & \frac{1}{8g^2T(R)}\int d^2\xt \Tr W^2 + \zt{H.c.} \nn
& = & \frac{1}{4g^2}\int d^2\xt \Tr W^2 + \zt{H.c.}.\ea
$T(R)$ denotes one half of the Dynkin index for the representation $R$, $\Tr(T^a T^b) = T(R) \xd^{ab}$.
The superfield $W_\xa \equiv W_\xa^a T^a$ in components is \ft{We mostly use the conventions in Wess and Bagger \cite{Wess:1992cp} including the choice of $\xs^\mu$ matrices and superderivatives. Our conventions differ only with regard to the normalization of the vector superfield and the integration of Grassmann variables.}
\[W_\xa = -\frac 1 8 \bar D^2 e^{-V} D_\xa e^V = -i (\xl_\xa + i\xt_\xa D +
\xt^\xb f_{\xa\xb} + i\xt \xt \CD_{\xa \dot \xa} \bar \xl^{\dot \xa}),\]
where $f_{\xa \xb}$ is the field strength in the spinor coordinate, $f_{\xa \xb} = -\frac 1 2 (\xs^\mu \bar \xs^\nu)_{\xa \xb} F_{\mu\nu} = -\xs^{\mu\nu}F_{\mu\nu}$.
The vector superfield $V$ in the Wess-Zumino gauge is,
$V = -2\xt^\xa \bar \xt^{\dot \xa} v_{\xa \dot \xa} + 2i \xt \xt \bar \xt
\bar \xl - 2i \bar \xt \bar \xt \xt \xl + \xt \xt \bar \xt \bar
\xt (D + i \p^\mu v_\mu)$. The integration over Grassmann numbers is defined by $\int \xt^2 d^2\xt = 2$.
Recall that the anomaly puzzle can be stated in terms of the absence of a supersymmetric anomaly equation. Such a possible equation is described by a supercurrent
$\CJ_\mu$, which is a superfield and can be defined as \cite{Shifman:1986zi}
\be\el{APN1SYM:1} \CJ_{\xa \dot \xa} \equiv -\frac{4}{g^2} \Tr[e^V W_\xa e^{-V} {\bar W}_{\dot \xa}] = -\frac{2}{g^2 T(R)} \Tr[e^V W_\xa e^{-V} {\bar W}_{\dot \xa}].\ee
Generally, the components of the supercurrent superfield are related to the $R$-current $R_\mu$, the supercurrent $J_{\xa\mu}$ and the stress tensor $\vartheta_{\mu\nu}$ respectively,
\ba\el{APN1SYM:supercurrent}
\CJ_{\xa\dot \xa} & = & C_{\xa\dot \xa} + \{\xt^\xb \chi_{\xb \xa \dot \xa} + H.c.\} + 2\xt^\xb \bar \xt^{\dot \xb} \tau_{\xa \dot \xa \xb \dot \xb} - \frac{1}{2} \{\xt_\xa \bar \xt_{\dot \xb} i \p^{\xg \dot \xb} C_{\xg\dot \xa} + H.c.\}\nn
& & + \{\frac 1 2 \xt^2 M_{\xa\dot \xa} + H.c\}  + \{\frac 1 2 \xt^2 \bar \xt^{\dot \xb} \bar \xl_{\dot \xb \xa \dot \xa} + H.c\} + \frac 1 4 \xt^2\bar
\xt^2 D_{\xa \dot \xa}.
\ea
$C_\mu, \chi_\mu$ and $\tau_{\mu\nu}$ are related to $R_\mu, J_\mu$ and $\xvt_{\mu\nu}$
as we shall see in the appendix.

For the supercurrent defined by \er{APN1SYM:1}, we have,
\be\el{MuSu2:rcurrent} C_{a\dot a} = R_{a\dot a} = -\frac{4}{g^2} \Tr(\xl_\xa \bar \xl_{\dot \xa}),\ee

The $\xt \bar \xt$ component
of $\CJ_{\xa \dot \xa}$, \eqref{APN1SYM:supercurrent} corresponds to the stress tensor and the exterior derivative of $R_\mu$. Note that $\tau_{\mu\nu}$ is not really the stress tensor as we shall see in the appendix. However,
the trace $\tau_\mu{}^\mu$ is proportional to that of the stress tensor $\xvt_\mu{}^\mu$.
So the operator $\tau_\mu{}^\mu$ also gives the trace anomaly.

As usual, this R-symmetry is broken at the quantum level because 
it is a chiral $U(1)$ symmetry.  The anomaly equation is,
\be\el{chiano}\p^\mu R_\mu = - \frac{T(G)}{16 \pi^2}F^a_{\mu\nu} \tilde F^{a\mu\nu},\ee
as follows from the Adler-Bardeen theorem \cite{Adler:1969er}. $T(G)$ is the $T(R)$ of the adjoint representation.

One can lift \er{chiano} to the supersymmetric form,
\be\el{APN1SYM:0a} \p_\mu \CJ^\mu = \frac i 2 {\cal C}\, \Tr [D^2 W^2 - \bar D^2 \bar W^2],\ee 
where $\cal C$ is some coefficient to be determined. The lowest component of \eqref{APN1SYM:0a} is the chiral anomaly equation, \er{chiano}. Equivalently,
we have,
\be\el{APN1SYM:0} \bar D^{\dot \xa} \CJ_{\xa \dot \xa} = {\cal C}\, D_\xa \Tr W^2.\ee

The real part of the $\xt$ component $\bar D^{\dot \xa} \CJ_{\xa \dot \xa}$ in \er{APN1SYM:0} corresponds to $\tau_\mu{}^\mu$ while the imaginary part corresponds to $-\p^\mu R_\mu$. This matches the $\xt$ component of $D_\xa \Tr W^2$, whose real and imaginary parts are $-\xe_{\xa\xb}\Tr (F F)$ and $-\xe_{\xa\xb}\Tr (F \tilde F)$ respectively.

To be consistent with the prediction of the Adler-Bardeen theorem, both sides should be bare operator and $\CC$ should be of one-loop just like \er{chiano}. However, to get the correct trace anomaly, $\xb$-function, which has higher-loop contributions should appear on the right hand side. So $\CC$ has to be proportional to $\xb$. Now we get the anomaly puzzle. At least, this is how this puzzle was
originally stated. There are quite a few subtleties as we shall see.

The situation becomes more complicated when matter
is introduced. The Lagrangian becomes,
\be\el{APN1SYM:Lmat} \CL = \frac{1}{4g^2}\int d^2\xt \Tr W^2 + \zt{H.c.} + \frac 1 4 \int d^4 \xt \sum_f \bar \Phi^f e^V \Phi_f.\ee
where $\Phi_f$ are chiral superfields describing matter. The supercurrent is defined as
\be \CJ_{\xa \dot \xa} = -\frac{4}{g^2} \Tr[e^V W_\xa e^{-V} {\bar W}_{\dot \xa}] + \frac 1 3 \sum_f \bar \Phi^f \lb \overleftarrow{\bar \nabla}_{\dot \xa} e^V \nabla_\xa - e^V \bar D_{\dot \xa}\nabla_\xa + \overleftarrow{\bar \nabla}_{\dot \xa}\overleftarrow{D}_\xa e^V\rb \Phi_f, \ee
where covariant derivative is introduced $\nabla_\xa \Phi_f = e^{-V} D_\xa e^{V} \Phi_f$. The $R$-current has the form,
\be \el{RcurrentFull}R_\mu = \frac 2 {g^2} \Tr (\xl \xs_\mu \bar \xl)- \frac 1 3 \sum_f
\lb \psi_f \xs_\mu \bar \psi_f - 2 i A_f \overleftrightarrow{\CD}_\mu
A_f^*\rb,\ee
where $A_f$ is the scalar component of the chiral superfield $\Phi_f$ and
$\psi_f$ is the spinor component.

With the introduction of matter, there is another $U(1)$
symmetry $\Phi_f \to e^{i\xa} \Phi_f$. The corresponding current is the so-called Konishi current (denoted by $R_\mu^f$). This symmetry is certainly chiral and its anomaly, the Konishi anomaly is given by,
\be\el{KonishiAn}\bar D^2 \CJ^f = \bar D^2 (\bar \Phi^f e^V \Phi_f) = \frac{T(R_f)}{2\pi^2} \Tr W^2.\ee

Equivalently, we can define a superfield $Q_{\xa \dot \xa}$ as in \er{Konishicurrent},
which has $R_\mu^f$ as its lowest component.

\section{Possible Solutions to the Puzzle}
Alternately, we can use Wilson effective action to describe the anomaly
puzzle \cite{Shifman:1986zi}. In this scenario, the theory has a large but finite cutoff. The Wilson effective action at scale $\xL$ is denoted by $S_W(\xL)$. Higher momentum modes can be integrated out to provide the running of the coupling constant. It can be shown that the new $S_W(\xL-\xd\xL)$ obtained by this renormalization group flow will only have a one-loop correction to the coefficient of $\Tr W^2$ in the Lagrangian \eqref{APN1SYM:Lmat}. This
agrees with the conclusion based on the non-renormalization theorem \cite{Seiberg:1993vc}. However, this result appears to be in contradiction to the multi-loop $\xb$
function. Note that the coefficient of the Wilson effective action (at scale $\xL$) can be related to the 1PI amplitude with an infrared cutoff $\xL$ (see \eg \cite{Kim:1998wz}). As noted by Shifman and Vainshtein, the absence of the infrared modes is the reason for the absence of multi-loop corrections in the $\xb$-function. Shifman and Vainshtein distinguish between the physical coupling constant and the corresponding coefficient in the Wilson effective action $S_W$. The latter is renormalized only at one-loop level as predicted by the nonrenormalization theorem. On the other hand, the physical coupling
can be obtained by evaluating the matrix elements (or the effective action). To do that, all the infrared modes have to be included and the higher-order corrections emerge.

Shifman and Vainshtein then proceed to propose that a single supercurrent
can contain both the stress tensor and the $R$-current. The anomaly equation for this bare supercurrent $\CJ_{\xa \dot \xa}$ is of the form of \er{APN1SYM:0} with a one-loop coefficient
$\CC$ (see \eg Eq.(19) in \cite{Shifman:1986zi}),
\ba\el{superanomalySV}
\bar D^{\dot \xa} \CJ_{\xa \dot \xa} & = & \frac 2 3 D_\xa\Big[ \frac {\xb_g^{(1)}(g_0)} {g_0^{3}} \Tr W^2 - \frac 1 8 \sum_f \xg_f \bar D^2 (\bar \Phi^f e^V \Phi_f)\Big]\nn
& = & \frac 2 3 D_\xa\Big[-\frac {3 T(G) - \sum_f T(R_f) } {16 \pi^2}  \Tr W^2 - \frac 1 8 \sum_f \xg_f \bar D^2 (\bar \Phi^f e^V \Phi_f) \Big]
\ea
The $\xb_g^{(1)}(g_0)$
is the one-loop $\xb$-function and $\xb_g^{(1)}(g_0)/g_0^3$ is a $g_0$-independent
number. Note that the operators in the equation are bare operators.
To obtain the physical coupling constant one needs to take the matrix elements of the operators on the right hand side. The matrix of element of $W^2$ is
shown to have finite multi-loop contribution that exactly reproduce the correct
$\xb$-function. More explicitly, when the operators on the right are expressed in terms of renormalized operators, \er{superanomalySV} becomes the anomaly equation that has the correct multi-loop $\xb$-function,
\be\el{APN1SYM:2} \bar D^{\dot \xa} \CJ_{\xa \dot \xa} = \frac 2 3 D_\xa \Big\{\frac {\xb_g(g)} {g^3} [\Tr\, W^2] - \frac 1 8 \sum_f \xg_f \bar D^2 [(\bar \Phi^f e^V \Phi_f)]\Big\},\ee
where $[\ ]$ indicates renormalized operators, and 
\be\el{betafunctionmatter} \xb_{g}(g) = - \frac {g3}{16 \pi^2} \frac {3 T(G)-\sum_f
T(R_f)(1-\xg_f)}{1- T(G) \xa/ 2\pi}.\ee
The term proportional to $\xg_f$ comes from the second term in \er{superanomalySV} because of the Konishi anomaly \eqref{KonishiAn}. More explicitly, the contribution (proportional to $\xg_f$) from the second term in \er{superanomalySV}
to the $\xb$-function follows when the operators on the right hand side are diagonalized.

Note that the trace of the stress tensor, $\xvt_\mu{}^\mu$ should be equal to $\sum \xb_a (M)O_a(M)$, in which $O_a (M)$ are renormalized operators
(at scale $M$). Moreover, \cite{Adler:1976zt} the coefficients of these operators $O_a(M)$ can be considered as $\xb$-functions (with $g_a(M)$ as variables) only when the operators $O_a (M)$ are ``orthonormal" at scale $M$. More explicitly, the operators $O_a (M)$ are chosen so that the corresponding matrix element
\ft{Without the insertion, this matrix element corresponds to a certain amplitude that defines the coupling constant at scale $M$.} of every coupling constant
$g_a$ only receives contribution from a single operator (on the right hand side of the trace anomaly) and the matrix element should be exactly unity (up to some power of $M$'s). Only in this case we can take the coefficients of the operators on the right hand side of the trace anomaly to be the $\xb$-functions.

However, there are some subtleties about \er{superanomalySV} \eml{that imply a contradiction with the Adler-Bardeen theorem.} Let us look at it more carefully. For the example of pure SYM, the $\xt^2$ component of an operator $W^2$ in fact has an imaginary part equal to (where $\xa_0 \equiv g^2_0/4\pi$),
\[-\frac 1 2 F\tilde F - \p_\mu (\xl \xs^\mu \bar \xl) = -\frac 1 2 F \tilde F - 4 \pi \xa_0 \p_\mu R^\mu,\]
where we used the fact that the second term on the left hand side is proportional to $\p^\mu R_\mu$ (see \er{MuSu2:rcurrent}).
After this term is moved to the left side in \er{superanomalySV}, it is clear that \er{superanomalySV} does not reproduce the Adler-Bardeen theorem; namely that the anomaly of the R-current $R_\mu$ is no longer proportional (with a coupling-constant-independent proportionality factor) to the topological term $F \tilde F$ as in the non-supersymmetry gauge theory. In other words, \er{chiano} no longer
holds as an operator equation of bare operators. Moreover, unlike it was previously claimed in the literature \cite{Jones:1983ip}, \er{APN1SYM:2} does not agree with the Adler-Bardeen theorem. Even if \er{APN1SYM:2} is not obtained from the one-loop equation (\ref{superanomalySV}) but is taken as the starting point, what appears on the right are just renormalized operators and can not be moved to the left side, which only contains bare operators. On the other hand, it has been shown \cite{Jones:1984mi} that if the lowest component of the supercurrent on the left of \er{superanomalySV} is taken as a renormalized operator, which is different from the bare $R$-current by a multiplicative renormalization factor, correct anomaly equations (for both trace anomaly and chiral anomaly) can be obtained.

There is another way to show the inconsistency between \er{superanomalySV} and the Adler-Bardeen theorem.
Together with the proposed expectation value of $W^2$ ((46) in \cite{Shifman:1986zi}),
\[\br W^2\ke = \lb 1 + \frac {T(G)\xa}{2\pi}+\dots \rb W^2_{\zt{ext}},\]
equation \eqref{superanomalySV} predicts a nonvanishing expectation value for the bare chiral current $\br \p^\mu R_\mu \ke$ at two-loop level. More explicitly, the two-loop value is $\frac {T(G)}{2\pi} \xa$ times
the one-loop value. This conclusion however, is in contradiction with the combination of the
Adler-Bardeen theorem and the proposed expectation value of $F \tilde F$ ((57) in \cite{Shifman:1986zi}),
\[\br F\tilde F \ke = (F\tilde F)_{\zt{ext}} \lb 1 + \frac {T(G)\xa}{\pi}\rb,\]
which implies a two-loop expectation value being $\frac {T(G)}{\pi} \xa$ times
the one-loop value. Note that the Adler-Bardeen theorem states that \er{chiano} is an operator equation of bare operators and the expectation values of both sides should have the same quantum corrections \ft{Higher order quantum corrections to $F\tilde F$ in QED are discussed in \cite{Anselm:1989gi}.}. Such an agreement is confirmed up to two-loop in \cite{Jones:1982zf}.

So eventually we have no choice but to construct two supercurrents. One of them, $\CJ^{(1)}{}_{\mu}$ has
the $R$-current as its lowest component, but does not have the stress tensor
among its components, while the other $\CJ^{(2)}{}_{\mu}$ has the stress tensor but not the $R$-current. As a result, there is no reason to have a single operator equation to describe both chiral anomaly and trace anomaly. The construction of two supercurrents using the background field method and dimensional reduction is first proposed by Grisaru et al \cite{Grisaru:1985yk}
\cite{Grisaru:1985ik} for the pure SYM and is further developed by Ensign and Mahanthappa \cite{Ensign:1987wy} for the coupled SYM. As we shall see later, there is an inconsistency in their calculation. However, we show by careful calculation that their results for the two currents are indeed correct.

Let us briefly review their results. In this approach, two different renormalized currents (both superfields) are defined. Each satisfies an anomaly equation with the anomaly expressed in terms of renormalized operators. One of the anomaly equations is similar to \er{APN1SYM:2} with the renormalized coupling constant and operators on the right hand side. The other has an one-loop coefficient for the $W^2$ term, which agrees with the Adler-Bardeen theorem
\ft{\eml{In \cite{Grisaru:1985ik} and \cite{Ensign:1987wy}, only the divergent contribution to the expectation value is considered and therefore the Adler-Bardeen
theorem implies the absence of any two-loop contribution to the coefficient $\CC$}.}. The explicit form is (Eq.(3.16) in \cite{Ensign:1987wy} after a change in convention),
\be\el{twosupABanomaly} \cd^{\xa \dot \xa} [J_{\xa \dot \xa} ] =  -\frac i 3 \frac {\xb_g^{(1)}} {g^3} \lb [\cd^\xa W^\xb \cd_\xa W_\xb] - [\bar \cd^{\dot \xa} \bar W^{\dot
\xb} \bar \cd_{\dot \xa} \bar W_{\dot \xb}]\rb.\ee
Both currents are renormalized operators whose expectation values are finite.

We shall see how these two supercurrents are constructed. For later convenience, we will give the operators that are involved,
\ba
W_{\xa \dot \xa} & = & \frac 4 {g^2_0}\Tr [e^{-V}\overline W_{\dot \xa} e^V
W_\xa] \\
K_{\xa \dot \xa} & = & \hhat W_{\xa \dot \xa} - \frac 1 {g^2} \Tr \lb i \hhat \xG_{\xb \dot \xa} \cd_\xa W^\xb - i \hhat \xG_{\xa \dot \xb} \bar \cd_{\dot \xa} \overline W^{\dot \xb} \rb\\
P_{\xa \dot \xa} & = & i (\bar \Phi e^V {\cd}_{\xa \dot \xa} \Phi - \bar \Phi {\overleftarrow {\cd}}_{\xa \dot \xa}  e^V \Phi)\\
\el{Konishicurrent} Q_{\xa \dot \xa} & = & -\frac 1 2 [\cd_\xa, \bar \cd_{\dot \xa}] (\bar \Phi e^V \Phi).
\ea
To be consistent with the expressions given above in section II, we use
the covariant derivatives in the gauge chiral representation. Note that $\hhat \xG_{\xa \dot \xa}$ introduced in \cite{Grisaru:1985ik} is the $\xe$-dimensional projection of the gauge connection and is gauge covariant under the $K$ gauge transformation (and invariant under the $\xL$ gauge transformation). Now in the gauge chiral representation, it is covariant under the $\xL$ gauge transformation. The expectation values of various operators are given by (3.3) (3.4) (3.5) in \cite{Ensign:1987wy}. They are obtained by the background field method and dimensional reduction. The dimension is $d = 4 -2 \xe$. Some of the expectation values given below are different from
from those in \cite{Ensign:1987wy}. This is due to different conventions.
For example, one of the equation given by (3.4) in \cite{Ensign:1987wy} is,
\be\el{twoloopQr} \br Q^r_{\xa \dot \xa}\ke^{(2)} =  0\times K_{\xa
\dot \xa}^e +\dots,\ee
The superscript $e$ denotes (renormalized) external fields. The superscript $r$ denotes renormalized fields ($\Phi = Z_\Phi^{1/2} \Phi^r$) with the field strength $Z_\Phi$
given by,
\be\el{fieldstrengthphi} Z_\Phi = 1 + 2\frac {g^2} \xe C(R) + g^4\lb \frac 1 {\xe^2} -\frac {1} \xe \rb(-3 T(G) C(R) + C(R) T(R) + 2 C(R)^2 ),\ee
and $Z_V$ given by
\[Z_V = 1+ \frac {g^2} \xe [3T(G) - T(R)] + \frac {g^4} \xe [3 T(G)^2 - T(G)T(R)-2C(R)T(R)].\]
However, $\br Q^r_{\xa \dot \xa}\ke^{(2)}$ are not the two-loop expectation values of the $Q^r_{\xa \dot \xa}$. Instead, it is the two-loop expectation value of the operator renormalized to one-loop order. In this paper, we use the symbol $\br \CO \ke^{(n)}$ for the expectation value of an operator $\CO$ without subtracting
any subdivergence due to renormalization of this operator. In this convention,
\er{twoloopQr} is expressed as 
\be\el{twoloopQ} \br Q^r_{\xa \dot \xa} + \frac {g^2} \xe (2C (R) Q^r_{\xa \dot \xa} + T (R) K^r_{\xa \dot \xa}) \ke^{(2)} = 0 \times K_{\xa \dot \xa}^e + \dots,\ee
where $C(R)$ is the quadratic Casimir operator of representation $R$. Note that only the two-loop contribution proportional to $K_{\xa \dot \xa}$ is evaluated
and the rest is unknown. With field strength renormalization $Z_\Phi$ ($Q_{\xa
\dot \xa} \equiv Z_\Phi Q_{\xa \dot \xa}^r$) given by \er{fieldstrengthphi}
and $\br K^r_{\xa \dot \xa} \ke^{(1)} = 0$, \er{twoloopQ} can be rewritten as,
\[\br Q_{\xa \dot \xa} \ke^{(2)} = 0 \times K_{\xa \dot \xa}^e + \dots.\]
Let us work with SQED, in which the corrections to the expectation value of $F \tilde F$ start at two-loop and the corrections to the right hand side of the Adler-Bardeen theorem start at three-loop. Naively, one can speculate that the $U(1)$ current in $Q_{\xa \dot \xa}$ satisfies the Adler-Bardeen theorem in the sense that there is no anomaly at two-loop level. Note that $\br Q_{\xa \dot \xa} \ke^{(1)}$ is nonvanishing and leads to a nonvanishing expectation value of $\p^\mu C_\mu^Q$ ($C_\mu^Q$ being the lowest component of $Q_\mu$).

However, $Q_{\xa \dot \xa}$ is not the correct superfield containing the anomalous $U(1)$ current. In the approach used by \cite{Ensign:1987wy}, $Q_\mu$ has to be renormalized
and the anomaly is described by a renormalized operator $[Q_\mu]$. However,
in the usual anomaly calculation of non-supersymmetric gauge theories (with
matter), the expectation value of
a bare chiral current $\p^\mu j^5_\mu$ is proportional to $F\tilde F$. Anyway, \eml{if we ignore this difference
and just apply the equations of motion on those bare fields from which $[Q_\mu]$
is constructed, the correct anomaly equation follows.}
 
Following Eq.(3.8) in \cite{Ensign:1987wy}, we can find out the relationship between the bare operators and the renormalized
operators as,
\ba \el{AnomalyN1SYM:1}
W_{\xa \dot \xa} & = &[W_{\xa \dot \xa}] - \frac {g^2} \xe (T(R) [W_{\xa \dot \xa}] -T(G) [K] - 3 C(R) [P_{\xa \dot \xa}] - C(R) [Q_{\xa \dot \xa}]) \nn
& & - \Big[\frac {g^4 } {4\xe} (3 T(G) T(R) + C(R) T(R)) + \frac {g^4}{\xe^2} (\frac 1 2 T(G) T(R) + \frac 1 2 C(R) T(R)) \Big][K_{\xa \dot \xa}] \nn
P_{\xa \dot \xa} & = & [P_{\xa \dot \xa}] - \frac {g^2} \xe (3 C(R) [P_{\xa \dot \xa}] + C(R)[Q_{\xa \dot \xa}] - T(R) [W_{\xa \dot \xa}]) \nn
& & + \Big[\frac {g^4 } {4\xe} (3 T(G) T(R) + C(R) T(R)) -  \frac {g^4}{2\xe^2} (T(G) T(R) - C(R) T(R)) \Big] [K_{\xa \dot \xa}]\nn
Q_{\xa \dot \xa} & = & [Q_{\xa \dot \xa}] - \frac {g^2} \xe T(R) [K_{\xa \dot \xa}]
\ea
Renormalized operators like $[W_{\xa \dot \xa}]$ are defined to have the expectation values of the background fields. With \er{AnomalyN1SYM:1}, we get to the conclusion that a current
\[\tilde \CJ_{\xa \dot \xa} \equiv W_{\xa \dot \xa} + P_{\xa \dot \xa} + \frac 1 3 Q_{\xa \dot \xa},\]
has no anomaly at two-loop level because of the lack of $g^4 [K_{\xa \dot \xa}]$ in $\br \tilde \CJ_{\xa \dot \xa} \ke$. However, this bare operator $\tilde \CJ_{\xa \dot \xa}$
contains neither $R_\mu$ nor $\xt_\mu{}^\mu$. Instead, two different renormalized operators need to be constructed to describe the two anomalies (trace and chiral). One of them is the supercurrent ($ \CJ^{(2)}{}_{\mu}$ in our notation),
\be \el{defsuprcurrent} \hat J_{\xa \dot \xa} \equiv \hat W_{\xa \dot \xa} -\frac 1 3 \hat K_{\xa \dot \xa}+ \hat P_{\xa \dot \xa} + \frac 1 3 \hat
Q_{\xa \dot \xa} + \CO(\xe),\ee
where the $\CO(\xe)$ terms do not affect
the $\xb$-function. The other is the Adler-Bardeen current ($\CJ^{(1)}{}_{\mu}$ in our notation),
\[[J_{\xa \dot \xa}] \equiv [W_{\xa \dot \xa}] + [P_{\xa \dot \xa}] + \frac 1 3 [Q_{\xa \dot \xa}].\]
As shown in \cite{Ensign:1987wy}, with the use of the equations of motion,
the desired anomaly equations \eqref{APN1SYM:2} and \eqref{twosupABanomaly} can be obtained.

A major problem this approach has is about the use of equation of motion. For example, the trace anomaly is described by \er{traceano}. The left hand side is a renormalized operator in the sense that expectation value of the operator $[\hat J_{\xa \dot \xa}]$ is given by the background
fields. There is no way this expectation value can give what appears on the
right hand side under a derivative $\bar \cd^{\dot \xa}$. In fact, the right hand side is obtained
by taking the expectation value of another operator $W^2$ which is obtained
by the equation of motion of $\hat K_{\xa \dot \xa}$. $\hat K_{\xa \dot \xa}$ is in the definition of the renormalized operator $[\hat \CJ_{\xa \dot \xa}]$ (see \er{defsuprcurrent}). The rest of $\hat \CJ_{\xa \dot \xa}$, the operator $\hat W_{\xa \dot \xa} + \hat P_{\xa \dot \xa} + \frac 1 3 \hat Q_{\xa \dot \xa}$ gives no anomalous contribution because of the EoM. This is very confusing. As shown in \cite{Ensign:1987wy},
the two sides of an equation of motion generally do not have the same expectation values. For example, $\bar \cd^{\dot \xa} W_{\xa \dot \xa} = 0$ (for pure SYM) follows from the equations of motion but apparently $\bar \cd^{\dot \xa} \br W_{\xa \dot \xa} \ke \ne 0$ following from \er{AnomalyN1SYM:1}
(see also Eq.(3.3) in \cite{Ensign:1987wy}). \eml{This problem will be further discussed in section V and a possible solution will be proposed.}

\section{Superpartner of the Trace Anomaly}
Before we move on to talk about the solution to the problem about the equation
of motion in the construction of the two supercurrents, let us give a supporting
argument for this approach. As explained in section III, the one-loop anomaly equation \er{superanomalySV}
implies that the operator $R_\mu'$, defined as the lowest component of $\hat
\CJ_\mu$, does not satisfy the Adler-Bardeen theorem. So it is unlikely that $R_\mu'$ is the the $R$-current $R_\mu$ \ft{Note that we define $R$-current as the $U(1)$ current associated with the (anomalous) symmetry that transforms the gaugino $\xl$, the matter scalar $A$ and the matter spinor $\psi$ according to the charge ratios of $1:\frac 2 3: - \frac 1 3$, which is determined by the classical supercurrent.}. Here we try to use some explicit calculation to show that for a general coupling $g$, this operator $R_\mu'$ is a mixing of the current $R_\mu$ and the Konishi current $R_\mu^f$ ($R_\mu^f$ being the $U(1)$ current in $Q_{\xa \dot \xa}$). This result clearly supports the approach
of two supercurrents. It also gives a clear physical interpretation
of the lowest component of $\hat J_{\xa \dot \xa}$ (the one containing $\xvt_{\mu\nu}$),
which is not given before. This validity of this interpretation is particularly clear at the infrared fixed point where the superconformal symmetry is restored. At this
point, the charges of those fields $\xl$, $A$ and $\psi$ under the $R$-symmetry are different from their classical values. This new $U(1)$ symmetry is also a classical symmetry whose current $R_\mu'$ is a linear combination of $R_\mu$ and the Konishi current $R_\mu^f$. The latter assigns charge $+1$ to both $A$ and $\psi$. Moreover, as we shall explain, this property of $R_\mu'$ agrees with the last term in the anomaly equation \er{superanomalySV}, which is actually not obtained in \cite{Ensign:1987wy}. So the $\CJ_{\xa \dot \xa}$
in the Shifman-Vainshtein scenario should be identified as the supercurrent
with $\xvt_{\mu\nu}$.
 
To study $R_\mu'$, we compute the Green's functions of this operator (or
rather $\p^\mu R_\mu'$) and various other fields. More explicitly, we compute the Green's functions with
an insertion of the operator obtained via supersymmetry transformation of the gamma trace (see below) of the supersymmetry current. This operator has a term $\p^\mu R_\mu'$ according to the superconformal algebra,
\be\el{supconfalgebra} \{S_{\xa},Q_{\xb}\} = 4 M_{\xa \xb} - 2 i D \xe_{\xa \xb} - 3 R' \xe_{\xa \xb}.\ee
where $M_{\xa\xb}$ are the Lorentz generators and $S_{\xa}$ is the generator corresponding to the gamma trace of $\bar J_\mu$. We will now show that the contact terms of the Green's function, which are the changes of the other fields under the transformation generated by $R_\mu'$, can be described by the transformation of a combination of the original $R$-symmetry and the Konishi $U(1)$ symmetry.

Let us start with the computation of $\xs^\mu \bar J_\mu$. The gamma trace of the supersymmetry current $\bar J_\mu$ in the Wess-Zumino model is,
\be
\el{gammatrace}
\xs^\mu \bar J_\mu =- i \xs^\mu (\bar \chi_\mu + \bar \xs_\mu \xs^\nu \bar \chi_\nu)
=  3i\xs^\mu \bar \chi_\mu = -2\sqrt 2\xs^\mu \p_\mu \bar \psi A \to -2\sqrt 2\xs^\mu \CD_\mu \bar \psi A
\ee
In the last step, we include the effect of the gauge field by covariantizing the derivative. \er{gammatrace} corresponds to the contribution from the matter multiplet to the gamma trace of the full supersymmetry current.

However, \er{gammatrace} does not vanish on-shell because of the interaction
with the gauge field. Let us consider SQED for simplicity. The equation of motion of $\bar \psi$ is,
\[i\CD_\mu\xs^\mu \bar \psi = - \sqrt 2i e A^* \xl.\]
So the gamma trace in 4d is
\[\xs^\mu \bar J_\mu = -2\sqrt 2\xs^\mu \CD_\mu \bar \psi A -4e AA^*\xl,\]
or equivalently, we should add $-\frac i 3 e\bar \xs_\mu \xl
A A^*$ to the definition of $\bar \chi_\mu$ in \er{gammatrace}. Of course, this term can also be obtained from explicit calculation (from a $ \xl \xs_\mu D$ term in the supercurrent
of the gauge multiplet). 
In $4+\xe$ dimension, the gamma trace becomes,
\ba
\xs^\mu \bar J_\mu & = & \xs^\mu \sqrt 2 \Big[\CD_\nu A \bar \xs^\nu \xs_\mu \bar \psi + \frac {4} 3 \bar \xs_{\mu \nu} \p^\nu (A \bar \psi)\Big] - i \xs^\mu [-\frac i 3 e\bar \xs_\mu \xl
A A^* + \bar \xs_\mu \xs^\nu (-\frac i 3 e\bar \xs_\mu \xl
A A^*)] \nn
& = & \sqrt 2\Big\{ (2+\xe)\CD_\mu A \xs^\mu \bar \psi+ \frac 1 3 \Big[-\frac
{(4+\xe)} 2 - (2+\xe) \Big]\p_\mu (A \xs^\mu \bar \psi) \Big\} -(4+ \xe)(1+\frac \xe 3) e AA^*\xl \nn
& = & \frac {\sqrt 2} 3 \xe \CD_\mu A\xs^\mu \bar \psi - \frac {2\sqrt 2}
3 \xe A\xs^\mu \CD_\mu \bar \psi -(\xe + \frac {4\xe} 3 )e AA^*\xl
\nn
& = & \frac {\sqrt 2} 3 \xe \CD_\mu A\xs^\mu \bar \psi -\xe\, e AA^*\xl.
\ea

The supersymmetry transformation (parametrized by $\xi_\xa$)  of the gamma trace is (note that the indices in $\bar \xs^\nu \xi$ are $(\bar \xs^\nu)^{\dot \xb \xb} \xi_\xb$),
\ba
\xd_\xi (\xs^\mu \bar J_\mu) & = & \xe\Big[-\frac 1 3 (\CD_\mu \psi \xs^\mu \bar \psi -2\sqrt{2} e A \bar \psi \bar \xl+ 2i \CD_\mu A^* \CD^\mu A)+\frac 1 {2\sqrt 2} e A^*\psi \xl \Big]\xe_{\xa \xb} \xi^\xb
\nn
& = & \Big[ \xe \sqrt 2 e  A \bar \psi \bar \xl+\xe\frac {\sqrt 2 e} {2} A^* \psi \xl + \dots\Big]\xe_{\xa \xb} \xi^\xb  \nn
\Rightarrow \Re [\xd_\xi (\xs^\mu \bar J_\mu)] & = & \Big[\xe\frac {3\sqrt 2 e} {4 } A^* \psi \xl + \zt{H.c.}\Big]\xe_{\xa \xb} \xi^\xb\ea
We use the equation of motion in the middle step. This is justified because \[(\CD_\mu \psi \xs^\mu + \sqrt 2 e A \bar \xl) \bar \psi,\]
is the the counting operator \cite{Lowenstein:1971jk}, which has a finite expectation value. In the last step, we just keep the
real part that is needed (and drop the imaginary term $\CD_\mu A^* \CD^\mu A$). Note that this is because the imaginary part should be proportional to the trace anomaly $\xvt_\mu{}^\mu$ following the superconformal algebra.

The Green's functions with an insertion of $\xd_\xi (\xs^\mu \bar J_\mu)$
can be evaluated. Alternately, the same results can be obtained by calculating the expectation value of
$\xd_\xi (\xs^\mu \bar J_\mu)$ in a certain background. The result bilinear in the external gaugino field $\xl$ is 
\ba \el{expxlxl}
\br -\xe {\sqrt 2} e A^* \psi \xl\ke_{\xl\bar\xl} & = & i \cdot (-i) \cdot \lb-{2e^2} \rb\xe\, \int \frac {d^4 p}{(2\pi)^4} \frac {\xl^e \pslash p \bar \xl^e}{p^2(p+k)^2} \nn
& = &- \frac {2ie^2}{(4\pi)^2} \xl^e \kslash k \bar \xl^e.
\ea
where $\xl^e$ is understood as Fourier transformation (with momentum suppressed)
of the external field $\xl^e(x)$. We have momentum $k$ flow into the vertex $-\xe {\sqrt 2} e A^* \psi \xl$
and for simplicity, we set the momentum exchange through the external field
$\xl^e$ to be $0$ and that through $\bar \xl^e$ to be $k$. Similarly we have scalar
contribution,   
\ba
\br -\xe {\sqrt 2} e A^* \psi \xl\ke_{AA^*} & = & i \cdot i \cdot \lb -2
e^2 \rb \xe\,\int \frac {d^4 p}{(2\pi)^4} \frac {\Tr[(\pslash p + \kslash k)\pslash p]}{p^2(p+k)^2}
A^e (A^e)^* \nn
& = & -\frac {4 i} {(4\pi)^2} e^2 k^2 A^e (A^e)^*.
\ea
Again, the momentum exchange through $(A^e)^*$ is set to be $0$ for simplicity.
Note that the expectation value proportional to $\psi^e \bar \psi^e$ is the same as \er{expxlxl} with the replacement of $\xl^e \to \psi^e$. Combine these results, the total expectation value is
\ba \el{mixRkonishi}
\br -\xe {\sqrt 2} e  A^* \psi \xl\ke & = & \frac {2ie^2}{(4\pi)^2} \xl^e \kslash k \bar \xl^e + \frac {2ie^2}{(4\pi)^2} \psi^e \kslash k \bar \psi^e
+\frac {4i} {(4\pi)^2} e^2 k^2 A^e (A^e)^* \nn
\el{expoverall} & = & -\frac {2ie^2}{(4\pi)^2}\Big[ (\xl^e \kslash k \bar \xl^e- \frac 1 3 \psi^e \kslash k \bar \psi^e + \frac 2 3 k^2 A^e (A^e)^*) + \frac 4 3 (\psi^e \kslash k \bar \psi^e + k^2 A^e (A^e)^*)\Big]
\ea
In the convention we are using, an operator with a momentum inflow $k$ is given by
\[\int \frac {d^4 x}{(2\pi)^4} e^{-i k \cdot x} O(x).\]
A current with the same charge $+1$ assigned to $\psi$ and $A$ (\ie Konishi current) is
\[i A \overleftrightarrow {\p} A^* + \psi \xs_\mu \bar \psi \to -i k^2 A A^* - i \psi \kslash k \bar \psi.\]
So it is clear that \er{expoverall} is a linear combination of $R_\mu$ and
the Konishi current $R_\mu^f$.

\er{mixRkonishi} is also consistent with \er{superanomalySV}, at
least loosely.
So the overall contribution to $\Re [\br \xd_\xi (\xs^\mu \bar J_\mu)\ke]$
is
\be\el{JconKonishi}\br\xe \frac {3e} {4\sqrt 2} A^* \psi \xl\ke \to- \frac {2 e^2}{(4\pi)^2} \p_\mu (\psi \xs^\mu \bar \psi).\ee
On the other hand, \er{superanomalySV} predicts a correction to the $U(1)$ current $R_\mu'$,
\[\frac {\xb_g^{(1)}(g_0)} {g_0^{3}} \Tr W^2 - \frac 1 8 \sum_f \xg_f \bar D^2 (\bar \Phi e^V \Phi) \to \xt \xt\Big[-\frac {\xb_g^{(1)}(g)} {g^{3}} \frac 1 2 F^2 + \frac i 4 \xg\p^\mu (\bar \psi \bar \xs_\mu \psi) + \dots \Big].\]
The first term is part of the trace anomaly $\xvt_\mu{}^\mu$. Compare this with the superconformal algebra \er{supconfalgebra},
we know that the correction to $R_\mu'$ is
\[-\frac \xg 2 \p^\mu (\bar \psi \bar \xs_\mu \psi) =- \frac {2e^{2}}{(4\pi)^2}
\p^\mu (\psi \xs_\mu \bar \psi).\]
which agrees with \er{JconKonishi}. So in some sense, we correctly calculate the $- \frac 1 8 \sum_f \xg_f \bar D^2 (\bar \Phi e^V \Phi)$ term, though
only in a way that is not manifestly supersymmetric.

However, our calculation is not without flaw. We use $- i \xs^\mu (\bar \chi_\mu + \bar \xs_\mu \xs^\nu \bar \chi_\nu)$ with $\bar \chi_\mu$ from $W_{\xa \dot \xa}$, the gauge field part of the supercurrent to get contribution
to the supersymmetry
current $\bar J_\mu$ from the gauge sector while the contribution from the
matter
sector is obtained from a direct generalization of \er{MuSu2:supcur} to $4+\xe$
dimension. Without this double standard, the coefficient in front of $\p_\mu (\psi \xs^\mu \bar \psi)$ in \er{JconKonishi} will be different and does
not match that predicted by \er{superanomalySV}. The physical
implication remains the same; namely that $R_\mu'$ receives correction proportional
to the Konishi current and is no longer the $R$-current that satisfies the
Adler-Bardeen theorem.  

It is not clear whether the operator defined by \er{RcurrentFull} generates a new $R$-symmetry at the fixed point. The terminology we use may be a little confusing. By saying ``new $R$-symmetry", we refer to the $U(1)$ symmetry that forms the superconformal algebra together with supersymmetry and the scaling, instead of the one that transforms the gaugino and matter fields according to the charge ratios of $1:\frac 2 3: - \frac 1 3$. The latter is referred to as $R$-symmetry. Anyway, it is not out of question that the operator of \er{RcurrentFull} can be the right current to generate the new $R$-symmetry. Note that at the fixed point, the trace $\xt_\mu{}^\mu$ scales various fields according to their ``quantum dimensions" instead of their canonical dimensions though the operator form of this dilatation current is defined according to the canonical dimensions.

In any explicit calculation, it is hard to see how the charges associated
with the operator defined by \er{RcurrentFull} can receive quantum corrections. So this operator, after some renormalization, is likely to be the $R_\mu$ that generates the anomalous chiral $U(1)$ symmetry as it is the case in QED. Anyway, in our opinion, the point is that there should be an anomalous current that transforms the fields according to the charge $1:\frac 2 3: - \frac 1 3$ and it satisfies the Adler-Bardeen theorem. Moreover, the latter is definitely not in the same multiplet as the stress tensor. The former may or may not be generated by the bare operator defined by \er{RcurrentFull} but this could depend on the calculation scheme and is hardly physically relevant.

\subsection{A Manifestly Supersymmetric Derivation}
The calculation above can be done using the dimensional
reduction and the background field method. In \cite{Ensign:1987wy}, the anomalous
dimension term is argued to be zero because of the assumption of on-shell external fields. The assumption of on-shell external fields is in general not justified and in this particular
case, leads to the missing of a term that
has physical meaning. In this subsection we recover the anomalous term so
that the super-anomaly equation of the current $\hat \CJ_{\xa \dot \xa}$ is exactly of the form of \er{APN1SYM:2}.

In this subsection, we are going to use the convention in \cite{Ensign:1987wy}.
So we need to determine the corresponding form (in this new convention) of
\er{APN1SYM:2}. According to \er{EMA28}, and Eq.(C39) in \cite{Ensign:1987wy},
\[\bar \cd^2 [\hhat \xG \cdot \hhat \xG] = - \xe W^\xa W_\xa.\]
we have the Konishi anomaly,
\be\el{Konishianomaly} \bar \cd^2 \br \bar \Phi \Phi \ke = - T(R)[W^2],\ee

Let $[\hat \CJ_{\xa \dot \xa}] $ be the supercurrent, renormalized so that its expectation value is finite and is exactly equal to what one would get by putting into $\hat \CJ_{\xa \dot \xa}$ the external fields alone \ft{Here, certain non-local contributions are ignored in the previous literature.}.  Note that $[\hat \CJ_{\xa \dot \xa}] $  is the supercurrent that contains the energy-momentum tensor in its  $\theta {\bar\theta}$ component .
Then the trace anomaly is the $\theta$ component of its super-trace, which
is given by Eq.(3.15) in \cite{Ensign:1987wy},
\be\el{traceano} \bar \cd^{\dot \xa} [\hat \CJ_{\xa \dot \xa}] = \frac 1 3 \frac {\xb_g} {g^3} \cd_\xa [W^2].\ee
The matter contribution to the $\xb$-function, up to two-loop, is (from \er{betafunctionmatter})
\[\frac {\xb_g} {g^3} = - 3 T(G) + T(R)(1-\xg) + \dots,\]
where $\xg$ is the anomalous dimension (defined from the anomalous scaling of the renormalized operators) The scaling dimension of the renormalized field $\Phi$ is $1 + \gamma/2$. 
In order for the missing (missing on the rhs of the anomaly equation \er{traceano})
term, $\bar \cd^2[\bar \Phi \Phi]$ to give the correct contribution to the $\xb$ function following \er{Konishianomaly} (see \cite{Shifman:1986zi}), we require \er{traceano} to be modified as,
\be\el{tanomwanod} \bar \cd^{\dot \xa} [\hat \CJ_{\xa \dot \xa}] = \frac 1 3 \lb\frac {\xb_g} {g^3} \cd_\xa
[W^2] + \xg \cd_\xa \bar \cd^2[\bar \Phi \Phi]\rb.\ee
Note that the numerical factors in this form are slightly different from
those in \er{APN1SYM:2}. Now both terms on the rhs follow from the vev of \[-\frac \xe {g^2} \cd_\xa W^2.\]
As explained in section III, the anomaly is determined by $\bar \cd^{\dot \xa} K_{\xa \dot \xa}$ and from
\[\bar \cd^{\dot \xa} K_{\xa \dot \xa} =  \frac \xe {g^2} \cd_\xa W^2 + \frac {3 \xe} 2  \bar \cd^2\cd_\xa
\bar \Phi e^V \Phi + 4 i \xe \bar \Phi e^{V} W_\xa \Phi,\]
we get $-\frac \xe {g^2} \cd_\xa W^2$. Let us now show the expectation value of the latter can indeed give
the correct super-anomaly equation \eqref{tanomwanod}. The contribution proportional to $\bar \cd^2 \bar \Phi^e \Phi^e$ in the vev of $W^2$ can be obtained in a similar
way as that of $W_{\xa \dot \xa}$. Now
instead of
\[W_{\xa\dot \xa} \to (\cd^2 \bar \cd_{\dot \xa} V) (\bar \cd^2 \cd_{\xa} V) + \dots,\]
we have the expansion,
\[W^2 \to \frac 1 2 (\bar \cd^2 \cd^{\xa} V) (\bar \cd^2 \cd_{\xa} V) + \dots,\]
as the vertex, where $V$ is the quantum fluctuation of the gauge field. The relevant diagrams are 1(d) (from two vertices of $\bar \Phi V \Phi$) and 1(e) (one vertex of $\frac 1 2 \bar \Phi V^2 \Phi$) in \cite{Ensign:1987wy}. It is not hard to see only the latter gives nonvanishing contribution. We have,
following the Feynman rule,
\[\br V V \ke = 2 g^2 \iv {\hat \Box},\]
that the vev of $W^2$ is
\ba
\br W^2 \ke_{\bar \Phi \Phi} = 1(e) & = & 2 g^2 \bar \Phi^e \iv \Box \overleftarrow{\cd}^\xa \overleftarrow {\bar \cd^2} \bar \cd^2 \cd_\xa \iv \Box \Phi^{e} \nn
& = & 4 g^2 \bar \cd^2 \bar \Phi^e \iv \Box \bar \cd^2 \cd^2 \iv \Box \Phi^{e} \nn
& = & \frac 4 \xe C(R) g^{2}\bar \cd^2 \bar \Phi^e \Phi^e \ea
In other words, the last term on the rhs of  \er{tanomwanod} is supposed to be
\be\el{vevW2Phi} \br -\frac {\xe} {g^2}\cd_\xa  W^2 \ke_{\bar \Phi \Phi}  = -4 C(R)g^2 \cd_\xa \bar \cd^2 \bar \Phi^e \Phi^e.\ee
The anomalous dimension $\xg$ is given in \cite{Shifman:1986zi} as,
\[\xg = - C(R) \frac \xa \pi \to - 4 C(R) g^2,\]
where we recall that $\alpha= g^2/(4\pi)$.
In the last step, we use the convention $(4\pi)^2 = 1$ in \cite{Ensign:1987wy}.
One can see that \er{vevW2Phi} exactly agrees with \er{tanomwanod}.

\subsection{Charges at the Infrared Fixed Point}
Previously, we have shown that the current $R_\mu'$ is a linear combination of the $R$-current $R_\mu$ and $R_\mu^f$. In this subsection, we apply our result to study the current $R_\mu'$ at the infrared fixed point of an
$SU(N)$ SYM that has $N_f$ matter fields $Q_f$ in the fundamental representation and
$N_f$ matter fields $\tilde Q_f$ in the anti-fundamental representation \ft{For a review on supersymmetric QCD and especially the properties at the infrared non-trivial fixed points, see \eg \cite{Intriligator:1995au}.}. The
current $R_\mu'$ is shown to be the anomaly-free
current, whose charge ratios for $\xl, A, \psi$ is,
\be\el{anomfreeratio} 1: \frac{N_f - N}{N_f}: -\frac{N}{N_f}.\ee
We then argue for the advantage of our method compared to the argument  \cite{Shifman:1999mv}
based on the approach in \cite{Shifman:1986zi}.

At the infrared fixed point, we have the current $R_\mu'$ as,
\[R_\mu' = R_\mu + \frac 1 3 \xg R_\mu^f.\]
This follows from the coefficient $3T(G) - \sum_f (1-\xg_f) T(R_f)$ in \er{APN1SYM:2} (and \er{betafunctionmatter}). For later convenience, let us rewrite \er{APN1SYM:2}
in the form,
\ba\el{APN1SYM:2d} \p^{\xa\dot \xa} \CJ_{\xa \dot \xa} & = -\frac i 3 D^2 \Big\{-\frac1 {16\pi^2}\Big[ \frac{3T(G)- \sum_f (1-\xg_f) T(R_f)}{1- T(G)\xa/2\pi}\Big] [\Tr\, W^2]\nn
& - \frac 1 8 \sum_f \xg_f \bar D^2 [(\bar \Phi^f e^V \Phi_f)]\Big\}.\ea
From the lhs of \er{APN1SYM:2d}, we have an operator $2\p^\mu R_\mu'$ \ft{In fact, we don't really need to know the factor in front of
$\p^\mu R_\mu'$.}. Taking its expectation value, the lowest order term is $2\p^\mu R_\mu^e$ (\ie \er{RcurrentFull} with all fields replaced by their
external counterparts). In the context of Slavnov-Taylor identity in the background field method \cite{Clark:1979te}, this term corresponds to the contact term and tells us the $R$-charges. Moreover, although the Adler-Bardeen theorem does not hold for this current $R_\mu'$, the lowest component of \er{APN1SYM:2d} still gives the chiral anomaly equation up to one-loop. So a connection between the factor in front of $\p^\mu R_\mu^e$ and the coefficient of $[\Tr W^2]$ can still be established. From \er{chiano} and the ratios of $3T(G)/\xg T(R_f)$ in \er{APN1SYM:2d}, we can infer that there is another renormalized operator $\frac 2 3 \xg [\p^\mu R_\mu^f]$ coming out of the term
$- \frac 1 8 \sum_f \xg_f \bar D^2 [(\bar \Phi^f e^V \Phi_f)]$. The reason
is that $R_\mu$ assigns charge $+1$ to $\xl$ while $R_\mu^f$ assigns charge $+1$ to $\psi_f$ and a combination of $R_\mu + \frac 1 3 \xg R_\mu^f$ gives the correct coefficient in the chiral anomaly equation (coefficient of $[\Tr W^2]$). Note that according
to \er{KonishiAn} the vev of the operator $\bar D^2 (\bar \Phi^f e^V \Phi_f)$ is going to give a term proportional to $\Tr[W^2]$ and a term proportional
to $\bar D^2 [(\bar \Phi^f e^V \Phi_f)]$. The $\xt \xt$ component of these two superfields have $f^2$ and $\p^\mu R_\mu^f$ respectively with the appropriate
coefficient determined by \er{chiano}.

Anyway, at the fixed point, there is no other contribution (from the anomaly)
of the form of $\p^{\xa \dot \xa} (\xl_\xa \bar \xl_{\dot \xa})$ \ft{As explained
above, the anomaly term $[\Tr W^2]$ has a contribution to the gaugino $U(1)$
current.} and we only have $\p^\mu R_\mu^e + \frac 1 3 \xg \p^\mu(R_\mu^f)^e$ (up to a factor) in the vev of $\bar D^{\dot \xa}\CJ_{\xa \dot \xa}$. As a result, the $R'$-charge at the fixed point for $\Phi$ is $\frac 2 3 (1 + \frac 1 2 \xg )$. At a different scales, the $R'$-charge is different. Naively this operator $R_\mu'$ behaves as a current that has different charges when it acts on states of different energy scales.

With anomalous dimension at the fixed point being,
\[\xg = 1 - \frac {3 N} {N_f},\]
which is necessary for the NSVZ $\xb$-function to vanish, the charge of $\Phi$
is,
\[\frac 2 3 (1 + \frac 1 2 \xg ) = \frac{N_f - N}{N_f}.\]
This result agrees with \er{anomfreeratio}.

The $R'$-charges at the fixed point can be obtained in
a very different way \cite{Shifman:1999mv}. In this case, a conserved current
is defined for every different $\xg$ (see Eq.(2.114) in \cite{Shifman:1999mv}),
\[\tilde \CJ_{\xa \dot \xa} \equiv \CJ_{\xa \dot \xa} - \frac {3T(G)- \sum_f (1-\xg_f) T(R_f)}{3 \sum_f T(R_f)} Q_{\xa \dot \xa},\]
At the infrared fixed point, the second term vanishes and this current is
just the supercurrent $\CJ_{\xa \dot \xa}$. However, the $R'$-charges are obtained from the form of $\tilde \CJ_{\xa \dot \xa}$ at the UV fixed point,
where $\xg_f = 0$. It is not clear why this works because $\tilde \CJ_{\xa
\dot \xa}$ with different $\xg_f$ are different operators. In other words,
$R'$-symmetries for different $\tilde \CJ_{\xa \dot \xa}$ are different. Why the charges of one symmetry is determined by that of the other needs to be explained. In our approach, the values of the charges for $R_\mu'$ come out naturally.

\section{Problems and Corresponding Solutions in the Approach Using two Supercurrents}
In this section, we justify the use of the equation of motion in \cite{Grisaru:1985ik} and \cite{Ensign:1987wy}. An equation of motion, when inserted
into the n-point functions serves like a functional derivative. For example, the expectation
value of the equation of motion of a field $\phi$, denoted as $\sS_{,\, \phi}$
satisfies
\[\br \sS_{,\, \phi}(x) X \ke = i\frac \xd {\xd \phi(x)} \br X \ke,\]
where $X$ denotes other operators. In the background field method, the expectation value of the equation of motion of $\phi$ gives something like
\be\el{trivialexp} i\frac {\xd \xG[\phi, \dots]} {\xd \phi(x)},\ee
where $\xG$ is the effective action.
\eml{In the standard non-supersymmetric calculation of the chiral anomaly using dimensional regularization, the chiral current is no longer conserved.
In other word, $\p^\mu j_\mu^5 \ne 0$ by the equation of motion. Instead, we have ($\psi$ being a Dirac spinor),}
\[\p^\mu j_\mu^5 = -\bar \psi \xg^5 \kslash D \psi + \zt{H.c.} + \frac 1
4 \bar \psi \{\overleftrightarrow{\kslash D}, \xg^5\} \psi\]
The first term on the right hand side and its Hermitian conjugate are proportional to the equation of motion and both vanish on-shell. The insertion of these terms in a Green's function only gives contact
terms (or in the context of expectation value, only trivial terms like \er{trivialexp})
but not any anomalous contributions.

However, in \cite{Grisaru:1985ik} and \cite{Ensign:1987wy} expectation values of the equations of motion apparently do have contributions from the anomaly terms.
This can be seen as follows.
From \er{AnomalyN1SYM:1}, we have 
\be\el{EoMnvanexp} \br \bar \cd^{\dot \xa} W_{\xa \dot \xa}\ke \sim T(G) W^2_{\zt{ext}} \ne 0.\ee
The right hand side is the anomalous contribution (given by the external
fields). Note that assuming no matter fields, the equation of motion of the gauge field implies, 
\be\el{EoMWnomatter} \bar \cd^{\dot \xa} W_{\xa \dot \xa} = 0.\ee
In \cite{Grisaru:1985ik} and \cite{Ensign:1987wy}, \eml{the lowest order contribution to the expectation value vanishes because of the
on-shell assumption} and a trivial expectation value (no anomaly) means a vanishing expectation value. Therefore, one expects
\be\el{EoMWnomatterexp} \br \bar \cd^{\dot \xa} W_{\xa \dot \xa} \ke = 0,\ee
which is in contradiction with \er{EoMnvanexp}. Moreover, when the equation of motion is used assuming that the expectation value of the equation of motion is trivial, which is necessary in the calculation of anomaly, the
nontrivial expectation value leads to an inconsistency. For example, \er{EoMWnomatter} is used on an operator 
\[\bar \cd^{\dot \xa} J_{\xa \dot \xa} = \bar \cd^{\dot \xa} \lb W_{\xa \dot \xa} - \frac 1 3 K_{\xa \dot \xa} \rb,\]
to get $-\frac \xe 3 \cd_\xa W^2$ (coming from the second term $-\frac 1 3
\bar
\cd^{\dot \xa} K_{\xa \dot \xa}$), which is then taken
to be the expectation value of $\bar \cd^{\dot \xa} J_{\xa \dot \xa}$. 
In other words, \er{EoMWnomatterexp} is assumed, which is apparently inconsistent
with \er{EoMnvanexp}.

In fact, in the scheme of dimensional reduction, the $U(1)_R$ current is actually conserved (by the equation of motion) and one may expect trivial expectation
values for $\p^\mu R_\mu$ and the supertrace of the supercurrent that has $R_\mu$ as its lowest component. This is possible only when we consider the non-local contributions to the expectation values.

In \cite{Grisaru:1985tc}, non-local contributions to the expectation values are considered but do not give any divergent contributions. The point is that it is necessary to include the non-local contributions in this calculation
scheme (background field method and dimensional reduction). For simplicity, the chiral anomaly in (non-supersymmetric)
QED is considered, as it was the same example that was discussed in \cite{Grisaru:1985ik}.  The non-local contribution to the expectation value of the chiral current $\bar \psi_{\dot \xa} \psi_\xa$ can be evaluated (in this calculation the superscript $e$ for the external field is dropped),
\ba
\br \bar \psi_{\dot \xa} \psi_\xa \ke & = &  i\cd^\xb{}_{\dot\xa} (\Box - i f)^{-1} {}_\xb{}^\xg \xe_{\xg\xa} \nn
& = &  i\cd^\xb{}_{\dot\xa} \iv \Box_0 (i\xe_{\xb \xg} A^{\mu}\p_{\mu} + f_{\xb\xg}) \iv \Box_0 (-iA^{\mu}\p_{\mu} \xd^{\xg}{}_{\xa}+i f^\xg{}_{\xa}) \iv \Box_0 \nn
& = & -ip{}^\xb{}_{\dot\xa} \iiint \frac 1 {p^2} f_{\xb\xg}(q) \frac 1 {(p-q)^2} f^\xg{}_{\xa}(k-q) \frac 1 {(p-k)^2} e^{-i k x} d^4 q d^4 p d^4 k\nn
& = & \int_0^1 2 dy \int_0^{1-y} dz \iint d^4 q d^4 k\, (-iyq -i z k){}^\xb{}_{\dot\xa} f_{\xb\xg}(q)f^\xg{}_{\xa}(k-q) \nn
& & \times \int d^4 l \frac 1 {(l^2 - \xD)^3} e^{-i k x}\nn
& = & - \frac i {2} \int_0^1 2 dy \int_0^{1-y} dz \iint d^4 q d^4 k\, \frac {(-i z k){}^\xb{}_{\dot\xa} f_{\xb\xg}(q)f^\xg{}_{\xa}(k-q)}{z^2 k^2 - z k^2} e^{-i k x}\nn
& = &  -\frac i 2 \iint d^4 q d^4 k\, \frac {i k_{\xa \dot\xa} f_{\xg\rho}(q)f^{\xg \rho}(k-q)}{k^2} e^{-i k x}
\ea
where $\xD = y^2 q^2 - yq^2 + z^2 k^2 - z k^2 + 2 y z q \cdot k$. In the
sixth line, we remove the $q$ dependence. Anyway, in the position space, the non-local contribution reads,
\[\br \p^{\xa \dot \xa} (\bar \psi_{\dot \xa} \psi_\xa) \ke = - \frac i 2 (f^2
- \bar f^2).\]
Note that the $A\p$ term, which we did not consider, has contribution too. But it appears that this contribution is proportional to $k_{\xa \dot \xa}F^2$, so after a procedure to make $\br \bar \psi_{\dot \xa} \psi_\xa\ke$ real, this contribution drops.

The non-local contribution is opposite to the contribution from the $\xe$-dimension operators, which is given in \cite{Grisaru:1985ik}. So an operator equation like,
\[\br \p^\mu j^5_\mu \ke = 0,\]
is valid. A renormalized operator $[j^5_\mu]$ defined in \cite{Grisaru:1985ik} has a nonvanishing expectation value that gives the correct chiral anomaly.

This result can be generalized to supersymmetric theories. In principle, one can compute the non-local contributions to the supercurrent $\CJ_\mu$ and show that its supertrace vanishes. Such contributions to some of the operators have actually been worked out in the literature. For example, the expectation value of the operator $\bar \Phi e^V \Phi$ has a (one-loop) non-local contribution (see \eg Eq.(6.7.10) in \cite{Gates:1983nr}). The contribution from the $\xe$-dimension operator can be found in Eq.(A28) in \cite{Ensign:1987wy},
\be\el{EMA28} \br \bar \Phi^r \Phi^r \ke^{(1)} = -2 C(R) \frac 1 \xe \bar \Phi^e \Phi^e - \frac 1 2 T(R) \frac 1 \xe \hhat \xG^e \cdot \hhat \xG^e,\ee
These two contributions are opposite to each other. So we have
\[ \bar D^2 \br \bar \Phi e^V \Phi \ke = 0.\]
On the other hand, a renormalized operator $[\bar \Phi e^{V} \Phi]$,
\[[\bar \Phi e^{V} \Phi] \equiv \bar \Phi e^{V} \Phi - \frac 1 \xe T(R) \hhat
\xG \cdot \hhat \xG.\]
can be defined to provide the correct Konishi anomaly \er{KonishiAn}. Here
we use a bare field $\Phi$ and therefore the first term on the right of \er{EMA28} in \cite{Ensign:1987wy} is removed.

The calculations of the non-local contribution that have be given in the literature
were not without flaw. The non-local contribution to $\br \bar \Phi e^V \Phi \ke$ is given by the same graphs that contributes
to the effective action (Eq.(8) in \cite{Grisaru:1985tc}, more explicitly $I_1,I_2,I_3$). In \cite{Gates:1983nr}, it appears that only part of the contribution ($I_3$) is considered. Moreover, $I_2$ is infrared divergent. It is not clear whether one can just drop this term.

However, non-local contributions to the expectation values of operators are infrared finite because the 4-momentum injected into the operators becomes an infrared cutoff. As shown by explicit calculation, the expectation value of operator $\bar \Phi e^V \Phi$ has terms containing $\cd^\xa W_\xa$
($W_\xa$ being the background field). These terms are finite and remain non-local after a differentiation by $\bar D^2$. We would like to see whether they have an impact on the anomaly. In \cite{Grisaru:1985ik}, they are discarded with the assumption of on-shell external fields.

Let us take a look at the $\xt \bar \xt$ component of $\cd^\xa W_\xa \cd^\xb W_\xb$. Note that all the fields are functions of $y^\mu = x^\mu + i \xt
\xs^\mu \bar \xt$. The expansion around
$x$ gives the $\xt^2 \bar \xt^2$ component. These two components ($\xt \bar \xt$, $\xt^2 \bar \xt^2$) are all that
are relevant in the chiral anomaly equation. First of all, let us take a
look at the bosonic contribution, which is of the form of $\p f^2$, from $\cd^\xa W_\xa \cd^\xb W_\xb$. To get
a nonvanishing $\xt \bar \xt$ component, we need one $\xt$ and one
$\bar \xt$. The lowest component of $\cd^\xa W_\xa$ vanishes identically.
Therefore both $\xt$ and $\bar \xt$ can not come from one of the $\cd^\xa W_\xa$. The $\xt$ (or $\bar \xt$) component of any scalar superfield like $\cd^\xa W_\xa$ is certainly fermionic. So there is no bosonic contribution
to the $\xt \bar \xt$ component (and the $\xt^2 \bar \xt^2$ component) of $\cd^\xa W_\xa \cd^\xb W_\xb$. A similar
argument can be made for $(\cd_\xa \cd^\xb W_\xb) W^\xa$.

In summary, we found that the terms containing $\cd^\xa W_\xa$ do not give contributions to the chiral anomaly. Nor do they give any
contributions proportional to $F^2$, which would appear in the trace anomaly. Note
that the derivative expansion does not apply to those $\cd^\xa W_\xa$ terms because of the infrared divergence in the expansion coefficients. As a result those $\cd^\xa W_\xa$ terms don't give local contributions even after being
acted on by $\bar D^2$. These terms have contributions quadratic in the spinor fields. It is not clear what their physical meanings are. Naively there can be such contributions
to the one-loop expectation value of the chiral current because of the Yukawa
coupling. Even though these contributions are just contact terms, the equation of motion is still spoiled. So in order to use the dimensional reduction, it appears that we have to assume that the external fields are on-shell, satisfying the classical equations of motion.

Note that other relevant contributions to the expectation value of $\bar \Phi e^V \Phi$ are proportional to $\cd^{(\xa} W^{\xb)} \cd_{(\xa} W_{\xb)}$. Let us also consider the $\xt \bar \xt$ component of $\cd^{(\xa} W^{\xb)} \cd_{(\xa} W_{\xb)}$, which is
\[\cd^{(\xa} W^{\xb)} \cd_{(\xa} W_{\xb)} \to f^{\xa \xb} i \bar \xt^{\dot \xa}\p_{\xa \dot \xa}(\xt^\xg
f_{\xg \xb}) \propto i\xt^\xa \bar \xt^{\dot \xa}\p_{\xa \dot \xa}f^2.\] So we have
\[\cd^{(\xa} W^{\xb)} \cd_{(\xa} W_{\xb)} + \zt{h.c} \to \xt^\xa \bar \xt^{\dot \xa}\p_{\xa \dot \xa} (if^2 - i\bar f^2).\]
With the assumption of $D^\xa W_\xa = 0$, $\cd^{(\xa} W^{\xb)} \cd_{(\xa} W_{\xb)}$ can be expressed as $D^2 W^2$ and we can get
to the usual form of the Konishi anomaly \er{KonishiAn}.

Despite these technical difficulties, it is quite tempting to expect similar non-local contributions to the expectation values of $W_{\xa \dot \xa}$, $P_{\xa \dot \xa}$ and $Q_{\xa \dot \xa}$. These non-local contributions cancel the contributions proportional to $K_{\xa \dot \xa}$ and make the use of the equations of motion justified.

\section{Rescaling Anomaly as an Infrared Effect}
The $\xb$-function obtained by the method introduced in \cite{Shifman:1986zi} agrees with the one obtained by a different method in \cite{ArkaniHamed:1997mj}. Moreover, the
role played by the field strength renormalization is almost the same. Take SQED as an example.
As pointed out in \cite{ArkaniHamed:1997mj} the rescaling of the field strength will change the coupling constant and the new theory (at cutoff $\xL$) is canonically normalized, i.e., has $Z = 1$. For
such a theory with $Z = 1$ both calculations predict that there is no quantum correction for process at the scale $\xL$. The effective Lagrangian after the rescaling is exactly the effective action for the external fields at the scale $\xL$.

The calculation (of the Jacobian) in \cite{ArkaniHamed:1997mj} involves UV
regularization and the infrared effects seem to be irrelevant. However, we
will show that if the momentum modes below an
arbitrary scale $\xL$ are ignored, there will be no contribution to the Jacobian
except for some non-renormalizable terms. The idea is to separate the contribution to the Jacobian from the modes above $\xL$. To do this, one can consider two Jacobians under the rescaling of field strengths
at cutoffs $\xL$ and $\xL'$ (assuming $\xL < \xL'$) respectively. The Jacobian
for the scaling of a chiral superfield $\Phi_f$ (in a representation of $R_f$) can be computed (Eq.(A.20) in \cite{ArkaniHamed:1997mj}),
\[\log J(\xL, e^{\xa(\xL)}) = -\frac 1 {16} \int d^2 \xt \frac {2 T(R_f)}{8\pi^2}\log (e^{\xa(\xL)}) W^2 + O(\frac 1 {\xL^4}).\]
Note that $J(\xL, e^{\xa(\xL)})$ can be understood as the Jacobian from the rescaling (by a factor of $e^{\xa(\xL)}$) of the momentum modes $k \le \xL$. We can set the scaling factor to be the same, then $J(\xL, e^{\xa(\xL')})$
is the contribution to $J(\xL', e^{\xa(\xL')})$ by modes $k \le \xL$. The difference $\log J(\xL', e^{\xa(\xL')})  - \log J(\xL, e^{\xa(\xL')})$ starts from the $\frac 1 {\xL^4} - \frac 1 {{\xL'}^4}$, which is the difference of the coefficient of a certain non-renormalizable term. \eml{This is consistent with the result about the contributions from the modes between $\xL$ and $\xL'$ in the Wilsonian renormalization group flow, in the sense that the effective theory with a cutoff $\xL$ generally has non-renormalizable terms proportional to the negative powers of $\xL$.}

The point is that the multi-loop contributions proportional to $\log \frac{\xL_0}{\xL'}$ do not come from the rescaling of modes between $\xL$ and $\xL'$. The choice of $\xL$ is arbitrary. In other words, the multi-loop contributions to the $\xb$-function do not come from any modes $k>\xL$. Therefore, the result in \cite{ArkaniHamed:1997mj} must also come from the infrared modes as in \cite{Shifman:1986zi}

\section{Conclusion}
In this paper, we have tried to elucidate and settle three problems that are related to the anomaly puzzle in $\CN = 1$ SYM. First, we study the properties of the current operator $R_\mu'$ that is in the same super-multiplet as the stress tensor. We show explicitly
that $R_\mu'$ is not the same as the (anomalous) current $R_\mu$ which transforms the fields according to the charge ratios $1:\frac 2 3: - \frac 1 3$. Only the anomaly of the latter current is of one loop order and satisfies the Adler-Bardeen theorem, while the anomaly of $R_\mu'$ is proportional to the $\xb$ function. By explicit calculation, we show that $R_\mu'$ is a mixing of the $R$-current, $R_\mu$, and the Konishi current. Moreover, we show that the $- \frac 1 8 \sum_f \xg_f \bar D^2 (\bar \Phi e^V \Phi)$ term that appears in the anomaly equation in \cite{Shifman:1986zi} gives the same mixed current $R_\mu'$
and therefore supports the existence of two different ``supercurrents," even though only one supercurrent was proposed in \cite{Shifman:1986zi}. 
 \eml{We then use supersymmetric
QCD at the infrared fixed point, as an example,
to show how the difference between $R_\mu'$ and $R_\mu$ can naturally be explained in terms of two-supercurrents.} 

Secondly, we show that non-local terms must be included for consistency when using the equations of motion in \cite{Grisaru:1985yk} and \cite{Ensign:1987wy}.  This is necessary because 
the equations of motion are used there with the assumption
that their expectation values trivially vanish, while they actually vanish only when the non-local contributions to the expectation values are included.

Finally, we compared the two different calculations of the NSVZ $\xb$ function in \cite{Shifman:1986zi} and \cite{ArkaniHamed:1997mj}.  The second method, which is based on the Jacobian arising from field strength rescaling, seems independent of the infrared behavior of the theory, while the first method seems to depend only on the infrared behavior. 
We resolve this apparent contradiction by showing that the infrared modes are also crucial in getting the multi-loop corrections to the $\xb$ function in the second method.  The reason, as we show, is that the contributions from modes above any arbitrary nonzero scale, $\xL$, to the rescaling Jacobian are proportional to non-renormalizable terms and therefore do not contribute to the $\xb$ function. 
 
\begin{acknowledgments}

We thank Steven Adler, Nathan Seiberg and Tim Jones for helpful comments. This work has been partially supported by NSF grants PHY-0071044 and PHY-0503366, and an RGI grant from the Graduate School of the University of Wisconsin-Milwaukee.

\end{acknowledgments} 
 
\appendix
\section{Supercurrent}
Here we will review some properties of the supercurrent. We mostly use the convention in \cite{Wess:1992cp} including the choice of $\xs^\mu$ matrices and superderivatives. The only two different choices are the form of vector superfield and the integration of Grassmann variables. First of all, let us put down the transformation rules for components of
a chiral field $\Phi$ that appears in the Wess-Zumino model with the Lagrangian,
\be\el{supercurrent:1} \CL = \frac 1 4 \int d^2 \xt d^2 \bar \xt \Phi \bar \Phi + \frac 1 2 \int d^2 \xt g \Phi^3.\ee
The chiral superfield can be written in the component form,
\be\el{MuSu2:chifield} \Phi = A + \sqrt 2 \xt \psi + \xt^2 F,\ee 
where the factor of $\sqrt 2$ is to make $\psi$ canonically normalized. The supersymmetry transformation (parameterized by $\xi, \bar \xi$) of the components is given by ($g = 0$ for simplicity),
\be \xd A = \sqrt 2 \xi \psi,\quad \xd A^* = \sqrt 2 \bar \xi \bar \psi,\ee
\be \xd \psi = i \sqrt 2 \xs^\mu \bar \xi \p_{\mu} A + \sqrt 2 \xi F,\quad \xd \bar \psi = -i \sqrt 2 \xi \xs^\mu \p_{\mu} A^* + \sqrt 2 \bar \xi F^*\ee
\be\el{MuSu2:susytran} \xd F = i \sqrt 2 \bar \xi \bar \xs^\mu \p_\mu \psi,\quad \xd F^* = i \sqrt 2 \xi \xs^\mu \p_\mu \bar \psi\ee
With this transformation, we can work out the current from Lagrangian,
\ba \el{MuSu2:Lag}\CL & = & i \p_n \bar \psi \bar \xs^n \psi + A^* \Box A + F^* F\nn
& = & \frac i 2 \p_\mu \bar \psi \bar \xs^\mu \psi - \frac i 2  \bar \psi \bar \xs^\mu \p_\mu \psi -\p_\mu A^* \p^\mu A + F^* F,
\ea
which is just \er{supercurrent:1} in component fields.

For simplicity, we set $F = F^* = 0$. Consider only the transformation generated by $\xi$, we have
\ba \el{MuSu2:current} \xd_\xi \CL & = &\sqrt 2\Big[ - \p_\mu A^* \p^\mu (\xi \psi) + \frac i 2 \p_\mu (- i \xi \xs^\nu \p_\nu A^*) \bar \xs^\mu \psi -\frac i 2 (- i \xi \xs^\nu \p_\nu A^*) \bar \xs_\mu \p^\mu \psi \Big]\nn
& = & \sqrt 2 \Big[- \p_\mu A^* \p^\mu (\xi \psi) + \frac 1 2(\p_\mu \xi )\xs^\nu \p_\nu A^* \bar \xs^\mu \psi+ \xi \xs^\nu \p_\mu \p_\nu A^* \bar \xs^\mu \psi - \frac 1 2 \xi \p_\mu (\xs^\nu \p_\nu A^* \bar \xs^\mu \psi) \Big]\nn
& = & \sqrt 2 \Big[- (\p^\mu \xi )\p_\mu A^* \psi -\xi \p_\mu (\p^\mu A^* \psi)+ \frac 1 2( \p_\mu \xi)\xs^\nu \p_\nu A^* \bar \xs^\mu \psi - \frac 1 2 \xi \p_\mu (\xs^\nu \p_\nu A^* \bar \xs^\mu \psi)\Big]\nonumber 
\ea
So the current is given by,
\be \el{MuSu2:current1} J_{\mu} = \sqrt 2 \p_\nu A^* \xs^\nu \bar \xs_\mu \psi. \ee
For a general theory, we need to keep the auxiliary fields when we do the variation. For a free theory, we can see that \er{MuSu2:Lag} implies no $F$ in $J$. The variation of the kinetic term $\bar \psi \xs^\mu\p_\mu \psi$ contains
something proportional to $\p_\mu (\xi F)$. The part with $(\p_\mu F)\xi$ is combined
with another term to form a total derivative $\p_\mu K^\mu$ but this $K^\mu$
does not appear in the conserved current since it has to be the same as the
part (from the variation of $\bar \psi \xs^\mu\p_\mu \psi$) proportional
to $(\p_\mu \xi) F$. 

Despite that, we can use equation of motion in the interacting theory to get $F$ in $J_\mu$, the current following Noether method does not contain $F$. All the interacting terms containing $F$ do not have derivatives and therefore cancel identically. The extra terms (proportional to coupling constant) in $J_\mu$ involves variation of $F$ though. 

Note that the charge generated by this current $Q \equiv \int J^0$ does generate
the correct supersymmetry transformation on component fields. This is obvious for
$\psi$ and $A$. One subtlety is the transformation on $\dot A$. It gives
\[\xd (\dot A) = -i\sqrt 2 \xi [\p_i A^* \xs^i \bar \xs_0 \psi, \dot A] =\sqrt 2 \xi \xs_0 \bar
\xs^i \p_i \psi = -\sqrt 2\xi \xs_0 \bar \xs^0 \p_t \psi = \sqrt 2\xi\dot \psi,\]
where we have used the equation of motion.


However, we actually use the $J_\mu$ given by, 
\be \el{MuSu2:supcur}J_\mu = \sqrt 2 \Big[\p_\nu A^* \xs^\nu \bar \xs_\mu \psi + \frac {4} 3 \xs_{\mu \nu} \p^\nu (A^* \psi)\Big].\ee
This is the ``improved" supersymmetry current of \cite{Ferrara:1974pz}, which gives the same charge and is also conserved.
Note that the second term does not have any contribution to the supersymmetry charge $Q$ because we have $\xs^{00} = 0$. Although $\xs^{0i} \ne 0$, the spatial derivatives do not contribute to the charge since they only give boundary terms when integrated over a spatial slice (volume). 

The supercurrent is given by,
\be \el{MuSu2:supcWZ} \CJ_\mu \equiv \frac 2 3 \Big[ i \Phi \stackrel {\leftrightarrow}{\p_\mu} \bar \Phi-\frac 1 4 D^\xa \Phi (\xs_{\mu})_{\xa \dot \xa} \bar D^{\dot \xa} \bar \Phi \Big]= \frac {2i} 3 \Phi \stackrel {\leftrightarrow}{\p_{\xa \dot \xa}} \bar \Phi+\frac 1 3 D_\xa \Phi  \bar D_{\dot \xa} \bar \Phi.\ee

Operator $\CJ_\mu$ \er{MuSu2:supcWZ} can certainly be expressed in the form of \er{APN1SYM:supercurrent}. Explicitly, we have
\be\el{MuSu2:rcurrentmatter} C_{\xa \dot \xa} =\frac {2i} 3 A {\overleftrightarrow \p}_{\xa \dot \xa} A^* + \frac 2 3 \psi_\xa \bar \psi_{\dot \xa}.\ee
This indicates a correct $2:-1$ charge ratios for bosonic and fermionic fields. The lowest component $C_\mu$ is related to R-current,
\[R_\mu = C_\mu.\]
By the way, with the introduction of both a mass term $m \Phi^2$ and a cubic term $g \Phi^3$, the $U(1)$ transformation
is no longer a symmetry. A charge given by the $R_\mu$ above does not give another supercharge $Q$ when we take their commutator $[R,Q]$.

The $\xt$ component $\chi_\mu$ of this current $\CJ_\mu$
is not
the supersymmetry current $J_\mu$. Instead, they are related by,
\be\el{MuSu2:supsymcur} J_\mu = i(\chi_\mu + \xs_\mu \bar \xs^\nu \chi_\nu).\ee
Now let us show this. Both second terms in \er{MuSu2:supcur} and \er{MuSu2:supsymcur}
give currents that are in the same equivalence class (second term on the right hand side) as \er{MuSu2:current1}. 

\ba \el{chicomp}\chi_\mu & = & \frac 2 3 \Big[i \sqrt 2 \psi \stackrel {\leftrightarrow}{\p_\mu} A^* - \frac i {\sqrt 2} \p_\nu A^* \xs^\nu \bar \xs_\mu \psi + \sqrt 2 \xe_{\xa \xb} F \bar \psi_{\dot \xa}\Big]\nn
& = & \frac 2 3 \Big[i \sqrt 2 \psi \stackrel {\leftrightarrow}{\p_\mu} A^* - \frac i {\sqrt 2} \p_\nu A^* \xs^\nu \bar \xs_\mu \psi - \frac {\sqrt 2} 2 \xs_\mu \bar \psi F \Big]\nn
& = & \frac {2\sqrt 2} 3\Big[- i \p_\mu \psi A^*- i (\frac 1 2 \p_\nu A^* \xs_\mu \bar \xs^\nu \psi + \p_\nu A^* \xs^\nu \bar \xs_\mu \psi)\Big].\ea
Note that we use $D = \p_\xt$, $\bar D = - \p_{\bar
\xt} -i 2\xt \p$. Now we have
\ba
J_\mu & = & i(\chi_\mu + \xs_\mu \bar \xs^\nu \chi_\nu)\nn
& = & \frac {2\sqrt 2} 3 i\Big[-(i \p_\mu \psi A^* + i \xs_\mu \bar \xs^\nu \p_\nu \psi A^*) - i (\frac 1 2 \p_\nu A^* \xs_\mu \bar \xs^\nu \psi + \p_\nu A^* \xs^\nu \bar \xs_\mu \psi)\nn
& & - i (\frac 1 2 \p_\nu A^* \xs_\mu \bar \xs^\xl\xs_\xl \bar \xs^\nu \psi + \p_\nu A^* \xs_\mu \bar \xs^\xl \xs^\nu \bar \xs_\xl \psi) \Big]\nn
& = & \sqrt 2 \Big[\p_\nu A^* \xs^\nu \bar \xs_\mu \psi +\frac 4 3 \xs_{\mu \nu} \p^\nu( \psi A^*)\Big], 
\ea
which is exactly \er{MuSu2:supcur}. In the third line, we use $\xs_\mu \bar \xs^\mu = -4$ and $\bar \xs_\nu \xs_\mu \bar \xs^\nu = 2 \xs_\mu$. 

The commutators of $Q$  with the component fields of $\Phi$ give the supersymmetry transformation of the fields when the equation 
of motion holds.

The $M_{\xa \dot \xa}$ for this supercurrent in fact vanishes. The $\xt^2$
component of $\Phi$ vanishes ($F = 0$) and therefore the first term in $\CJ_\mu$ does not give any contribution. On the other hand, $D\Phi$ has neither
 $\xt^2$ nor $\bar \xt^2$ component. So to get a $\xt^2$, we need a $\xt$
 from each factor ($D\Phi$ and $\bar D \bar \Phi$), $D\Phi$ only has a nonvanishing
 $\bar \xt$\ component. The $\xt$ component is just $F$ and therefore vanishes.
 Similarly, the $\bar \xt$ component of $\bar D \Phi$ vanishes and we have
 a vanishing $\bar M_\mu$.

The bosonic part of the $\xt \bar \xt$ component of the supercurrent can be worked out following from \er{MuSu2:supcWZ},
\be\el{xtbarxtcom}
\frac 2 3 \Big[2\p_{\xa\dot \xb}A\, \bar \xt^{\dot \xb} \xt^\xb\p_{\xb \dot \xa} A^* -  \xt^\xb \bar \xt^{\dot \xb} (\p_{\xb \dot \xb} A \stackrel{\leftrightarrow}{\p_{\xa
\dot \xa}} A^* - A \stackrel{\leftrightarrow}{\p_{\xa\dot \xa}}\p_{\xb \dot \xb} A^*)\Big]
\ee

We use $v_{\mu\nu}$ to denote the $\xt \bar \xt$ component and $t_{\mu\nu}
\equiv \frac 1 2 v_{(\mu\nu)}$ as one half of the symmetric part of the $\xt \bar \xt$ component. With a $t_{\mu\nu}$ defined as
\be \el{supercurtbos} t_{\mu\nu} = \frac 2 3 \Big[-2\p_{(\mu}A \p_{\nu)} A^* +\frac 1 2 \eta_{\mu\nu} \p_\rho A \p^\rho A^*+ \frac 1 2 \p_\mu \p_\nu A\, A^* + \frac 1 2 A\p_\mu \p_\nu A^*\Big],\ee
we can obtain the $\xt \bar \xt$ component explicitly,
\ba
& & 2 t_{\xa \dot \xa \xb \dot \xb} + \frac i 2 \xe_{\xa\xb} \xe_{\dot
\xb \dot \xg} (\p^{\xg \dot \xg} R_{\xg \dot \xa}) -\frac i 2 \xe_{\xb \xg} \xe_{\dot \xa \dot \xb}  \p^{\xg\dot \xg} R_{\xa \dot \xg}\nn
& = & \frac 2 3 \Big[-\p_{\xa\dot \xa} A\ \p_{\xb \dot \xb} A^* - \p_{\xb \dot \xb} A\ \p_{\xa \dot \xa} A^*  - 2 \p_{\xa \dot \xb} A \p_{\xb \dot \xa} A^*  + \p_{\xa \dot \xa} \p_{\xb \dot \xb} A\, A^* + A\p_{\xa \dot \xa} \p_{\xb \dot \xb} A^*\Big].
\ea
Note that this agrees with \er{xtbarxtcom}.

Let take a look at the relationship between $t_{\mu\nu}$ and the stress tensor.
The stress tensor can be obtained as,
\ba
T_{\mu\nu} & = & - \p_{\mu} A \p_{\nu} A^* - \p_{\nu} A \p_{\mu} A^* + \eta_{\mu\nu} \p_a A \p^a A^* -\frac i 4 (\bar \psi \bar \xs_\mu \p_\nu \psi + \bar \psi \bar \xs_\nu \p_\mu \psi) \nn 
& &  - \frac i 4 (\psi \xs_\mu \p_\nu \bar \psi + \psi \xs_\nu \p_\mu \bar \psi) + \eta_{\mu\nu} (\frac i 2 \bar \psi \bar \xs^\rho \p_\rho \psi
+ \frac i 2 \psi \xs^\rho \p_\rho \bar \psi).
\ea

Of course, we usually use $\vartheta_{\mu\nu}$, which is defined as the improved stress tensor of $T_{\mu\nu}$,
\ba
\vartheta_{\mu\nu} & = & T_{\mu\nu} + \frac 1 3 (\p_\mu\p_\nu - \eta_{\mu\nu}\Box)
AA^*\nn
& = & -\frac 2 3 \p_{\mu}A \p_{\nu} A^* -\frac 2 3 \p_{\nu}A \p_{\mu} A^*+\frac
1 3 \eta_{\mu\nu} \p_{\rho}A \p_{\rho} A^* \nn
& & + \frac 1 3 \p_\mu \p_\nu A\, A^* + \frac 1 3 A\p_\mu \p_\nu A^* - \frac 1 3 \eta_{\mu\nu} (A\Box A^*+\Box
A\, A^*).
\ea
The extra term to $T_{\mu\nu}$ does not provide any contribution to the charges $P_\mu$ and certainly does not affect the conservation. Note that we have $t_\rho{}^\rho = \frac 1 3 (A\Box A^*+\Box A\, A^*)$, and therefore
\ba
\el{tcompandvarxt}
t_{\mu\nu} - \eta_{\mu\nu} t_\rho{}^\rho & = & -\frac 4 3 \p_{(\mu}A \p_{\nu)} A^* +\frac 1 3 \eta_{\mu\nu} \p_\rho A \p^\rho A^*+ \frac 1 3 \p_\mu \p_\nu A\, A^* + \frac 1 3 A\p_\mu \p_\nu A^*  \nn
& & - \frac 1 3 \eta_{\mu\nu} (A\Box A^*+\Box A\, A^*)\nn
& = & \vartheta_{\mu\nu}.
\ea
Of course, following \er{tcompandvarxt}, we have $\vartheta_\mu{}^\mu$ proportional to $A\Box A^* + \Box A\, A^*$.
Similar conclusion should hold for improved supersymmetry current $J_\mu$.

Note that this difference between $t_{\mu\nu}$ and $\vartheta_{\mu\nu}$
follows from the difference between $\chi_\mu$ and $J_\mu$ \er{MuSu2:supsymcur}.
This statement is actually only true when $t_{\mu\nu}$ is symmetric. With \er{tcompandvarxt} the variation of $J_\mu$ (under a supersymmetry transformation
generated by $\bar \xi$) is proportional to $\vartheta_{\mu\nu}$ while that of $\chi_{\mu}$ is proportional to $t_{\mu\nu}$.

Let us now include the fermionic fields in our considerations. The $\xt \bar \xt$ component, $v_{\mu\nu} (\xt \xs^\nu \bar \xt) $, of the supercurrent is (following from \er{MuSu2:supcWZ}),
\ba \el{vcompWZ}
v_{\mu\nu} (\xt \xs^\nu \bar \xt) & = & \frac {4 i} 3 (\xt \psi) \stackrel{\leftrightarrow}{\p_\mu} (\bar \xt \bar \psi) - \frac 1 3 \psi(y) \xs_\mu \bar \psi (y^+) \nn
& & -\frac 1 6 \Big[-i 2 \cdot 2 (\bar \xt \xs^\nu)^\xa \p_\nu (\xt \psi) (\xs_\mu \bar \psi)_\xa + i 2 \cdot 2 (\psi \xs_\mu)_{\dot \xa} (\xt \xs^\nu)^{\dot \xa} \p_\nu (\bar \xt \bar \psi)\Big]\nn
& = & (\xt \xs^\nu \bar \xt)\frac 2 3 \Big[-i \psi \xs_\nu \stackrel{\leftrightarrow}{\p_\mu}\bar \psi - \frac i 2  \p_\nu \psi \xs_\mu \bar \psi + \frac i 2 \psi \xs_\mu \p_\nu \bar \psi - \frac i 2 \p_\rho \psi \xs_\nu \bar \xs^\rho \xs_\mu \bar \psi - \frac i 2 \p_\rho \bar \psi \bar \xs_\nu \xs^\rho \bar \xs_\mu \psi\Big] \nn
& = & (\xt \xs^\nu \bar \xt) \frac 2 3 \Big\{ -i \psi \xs_\nu \stackrel{\leftrightarrow}{\p_\mu}\bar \psi - \frac i 2  \p_\nu \psi \xs_\mu \bar \psi + \frac i 2 \psi \xs_\mu \p_\nu \bar \psi \nn
& & +\Big[ \frac 1 2  \xe_{\nu \rho \mu a} \p^\rho \psi \xs^a \bar \psi - \frac i 2 \p_\rho \psi (\eta_{\nu \mu} \bar \xs^\rho - \xd^\rho_\mu \bar \xs_\nu - \xd^\rho_\nu \bar \xs_\mu) \bar \psi + \zt{H.c.}\Big]\Big\}
\ea

We would like to show that the anti-symmetric part
of $v_{\mu\nu}$ satisfies the following expression on-shell,
\be\el{antisymv} v_{[\mu\nu]} = \frac 1 2 \xe_{\mu\nu a b} \p^a R^b .\ee

It is not hard to rewrite $\frac 1 2 \xe_{\mu\nu a b} \p^a R^b$ in the form given by,
\be\el{antiRdspinor} \frac 1 2 \xe_{\mu\nu a b} \p^a R^b = \frac i 2 \xe_{\xa \xb} \p^\xg {}_{(\dot \xb} R_{\xg \dot \xa)} + \zt{H.c.}\ee
Following, 
\[\xs^{[\mu}_{\xa \dot \xa} \xs^{\nu]}_{\xb \dot \xb} = (\xs^{\mu\nu}\xe)_{\xa
\xb} \xe_{\dot \xa \dot \xb} + (\xe\bar \xs^{\mu\nu})_{\dot \xa
\dot \xb} \xe_{\xa \xb}.\]
we have
\ba
-\frac 1 2 \xs^\mu_{\xa \dot \xa} \xs^\nu_{\xb \dot \xb} \xe_{\mu\nu a b} \p^a R^b & = & -\frac 1 2 [(\xs^{\mu\nu}\xe)_{\xa
\xb} \xe_{\dot \xa \dot \xb} + (\xe\bar \xs^{\mu\nu})_{\dot \xa
\dot \xb} \xe_{\xa \xb}] \xe_{\mu\nu a b} \p^a R^b \nn
& = & [i(\xs^{a b}\xe)_{\xa \xb} \xe_{\dot \xa \dot \xb} - i (\xe\bar \xs^{ab})_{\dot \xa \dot \xb} \xe_{\xa \xb}]  \p_a R_b \nn
& = &\frac 1 2 [i \p_{(\xa \dot \xg} R^{\xg \dot \xg} \xe_{\xg \xb)} \xe_{\dot \xa \dot \xb} - i \p^{\xg} {}_{(\dot \xa} R_{\xg \dot \xb)}\xe_{\xa \xb}].
\ea

Now we have
\ba \el{antisymvspin}\frac 1 2 \xe_{\xa \xb} v_{\mu\nu}\xs^\mu_{\xg (\dot \xa} \xs^\nu_{\xd \dot \xb)}\xe^{\xd \xg} & = & i\xe_{\xa \xb} \psi_\xd \p_{\xg (\dot \xa} \bar \psi_{\dot \xb)} \xe^{\xd \xg}  - i \p^{\xg} {}_{(\dot \xa} R_{\xg \dot \xb)}\xe_{\xa \xb}\nn
& = & i \xe_{\xa \xb} \p^\xg{}_{(\dot \xa} (\psi_\xg \bar \psi_{\dot \xb)}) - i \p^{\xg} {}_{(\dot \xa} R_{\xg \dot \xb)}\xe_{\xa \xb}\nn
& = & \frac 1 2 i \xe_{\xa \xb} \p^\xg{}_{(\dot \xa} R_{\xg \dot \xb)}.
\ea
This is the part antisymmetric in $\xa,\xb$ and symmetric in $\dot \xa, \dot \xb$ and corresponds to the self-imaginary dual component of the $\tau_{[\mu\nu]}$.
The Hermitian conjugate gives another term,
\[-\frac 1 2 i \xe_{\dot \xa \dot \xb} \p_{(\xa}{}^{\dot \xg} R_{\xb) \dot \xg}\]
With the use of \er{antisymvspin} and \er{antiRdspinor}, we get \er{antisymv}.

Note that in the derivation above, we did use equation of motion. In fact, \er{antisymv}
is true on-shell. It follows from 
\be\el{conservationsuperc} D^\xa \CJ_{\xa \dot \xa} = 0,\ee
which can be derived using the explicit form \er{MuSu2:supcWZ} and equation of motion (and its conjugate)
\[D^2 \Phi = 0,\]
and,
\[[\bar D_{\dot \xa}, D^2 ] = 4 i D^\xa \xs^\mu_{\xa \dot \xa} \p_\mu.\]

In component (more explicitly 
$\bar \xt$ component), \er{conservationsuperc} implies
\ba i (\xs^\nu \bar \xt)^\xa \p_\nu R_{\xa \dot \xa} + (\xs^\nu\bar \xt)^\xa v_{\xa \dot \xa \nu} & = & i\p_\nu R_\mu \bar \xt \xe \bar \xs^\nu \xs^\mu + v_{\mu\nu} \bar \xt \xe \bar \xs^\nu \xs^\mu \nn
& =& \p_a R_b \xe^{a b \mu \nu} \bar \xt (\xe \bar \xs_{\mu \nu}) +2 v_{[\mu\nu]}
\bar \xt (\xe \bar \xs^{\nu\mu}) = 0.
\ea
This is equivalent to \[v_{[\mu\nu]} = \frac 1 2 \xe^{\mu \nu a b } \p_a R_b,\]
which is exactly \er{antisymv}.

Let us now consider SQED, whose supercurrent is given by 
\[\CJ_{\xa \dot \xa} \equiv -\frac 2 {e^2} W_\xa \bar W_{\dot \xa}.\]
We have $R_{\xa \dot \xa} = -\frac 2 {e^2}\xl_\xa \bar \xl_{\dot \xa}$. The $\xt \bar \xt$ component is decomposed in a way as \er{APN1SYM:supercurrent}
(with only fermionic fields considered),
\ba \el{xtbarxtdecom}2\xt^\xb \bar \xt^{\dot \xb} \frac 1 {e^2}(i \xl_\xa \p_{\xb \dot \xb}\bar \xl_{\dot \xa} - i \p_{\xb \dot \xb} \xl_\xa \bar \xl_{\dot \xa}) & = & 2\xt^\xb \bar \xt^{\dot \xb} \frac 1 {e^2}(i \xl_{(\xa} \p_{\xb) \dot \xb} \bar \xl_{\dot \xa} - i \p_{\xb (\dot \xb} \xl_\xa \bar \xl_{\dot \xa)})\nn
& & +2 \xt^\xb \bar \xt^{\dot \xb} \frac 1 {e^2}(\frac i 2 \xe_{\xa \xb} \xl^\xg \p_{\xg \dot \xb}\bar \xl_{\dot \xa} - \frac i 2 \xe_{\dot \xa \dot \xb} \p_{\xb \dot \xg} \xl_\xa \bar \xl^{\dot \xg}) \nn
& \equiv & 2\xt^\xb \bar \xt^{\dot \xb} \tau_{\xa \dot \xa \xb \dot \xb} + \frac 1 2 \xt_\xa \bar \xt_{\dot \xb} ( i \p^{\xg \dot \xb} R_{\xg \dot \xa} + \zt{H.c}),\ea
where $\tau_{\mu\nu}$ is given by,
\ba\el{supercurvcom} 
\tau_{\mu\nu} & = &  \frac 1 4 \bar \xs_\mu^{\dot \xa \xa} \bar \xs_\nu^{\dot \xb \xb} \frac 1 {e^2}\Big[\frac 1 2 ( i \xl_\xa \p_{\xb \dot \xb} \bar \xl_{\dot \xa} + i \xl_\xb \p_{\xa \dot \xb} \bar \xl_{\dot \xa}  - i \p_{\xb \dot \xb} \xl_\xa \bar \xl_{\dot \xa} - i \p_{\xb \dot \xa} \xl_\xa \bar \xl_{\dot \xb} )
\nn 
& & - (\frac i 2 \xe_{\xa \xb} \p_{\xg \dot \xb} \xl^\xg \bar \xl_{\dot \xa} + \zt{H.c})
\Big]\nn
& = &  -\frac 1 {4 e^2}( i\xl \p_\nu \xs_\mu \bar \xl - \frac i 2 \xl \xs_\nu \bar \xs^\rho \xs_\mu \p_\rho \bar \xl +\zt{H.c}) -\frac 1 {e^2} (\frac i 8 \p_\rho \xl \xs^\rho \bar \xs_\nu \xs_\mu \bar \xl + \zt{H.c}) \nn
& = & -\frac 1 {e^2}(\frac i 4 \xl \p_\nu \xs_\mu \bar \xl - \frac i 8  \xl \xs_\nu \bar \xs^\rho \xs_\mu \p_\rho \bar \xl) - \frac 1 {e^2}(\frac i 8 \xl \xs_\mu \bar \xs^\rho
\xs_\nu \p_\rho \bar \xl + \frac i 4 \xl \xs_\mu \p_\nu \bar \xl) + \zt{H.c}\nn
& = & -\frac 1 {e^2}(\frac i 2 \xl \p_\nu \xs_\mu \bar \xl +\zt{H.c})-\frac 1 4 \xe_{\mu\nu b a} \p^b (\frac 1 {e^2}\xl \xs^a \bar \xl).
\ea
Note that we have a $-\frac 1 4 \xe_{\mu\nu a b} \p^a R^b$ in $\tau_{\mu\nu}$. So $\tau_{\mu\nu}$
is not symmetric. Following a similar derivation as in the Wess-Zumino model, we can show that the $\tau_{[\mu\nu]} = 0$ and certainly, the symmetric part produces the stress
tensor. This symmetric part $t_{\mu\nu} \equiv \tau_{(\mu\nu)}$ is in fact
related to the stress tensor in the same way as \er{tcompandvarxt},
\be\el{xvtt} \xvt_{\mu\nu} = t_{\mu\nu} - \eta_{\mu\nu} t_\xl{}^\xl.\ee
 
\goodbreak

\end{document}